\documentclass{WileyMSP-template}
\usepackage{amsmath}
\usepackage{amssymb}
\usepackage{mathtools}
\usepackage{xcolor}
\usepackage{bm}
\usepackage{appendix}
\usepackage{braket}
\usepackage{hyperref}

\newcommand{\U}[1]{_{#1}}
\newcommand{\Ul}[1]{\underline{#1}}
\newcommand{\Ho}[1]{^{\mathrm{#1}}}

\newcommand{\dt}[0]{\frac{\partial}{\partial t}}
\newcommand{\DT}[0]{\frac{\mathrm{d}}{\mathrm{d} t}}
\newcommand{\dz}[0]{\frac{\mathrm{d}}{\mathrm{d} z}}

\newcommand{\bs}[1]{\boldsymbol{#1}}

\begin{document}
\pagestyle{fancy}

\rhead{\includegraphics[width=2.5cm]{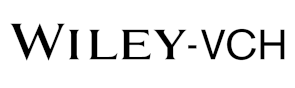}}


\title{Electronic correlation effects in the response of graphene and MoS$_2$ monolayers to the impact of highly-charged ions}

\maketitle

\author{Giorgio Lovato\textsuperscript{1}*}
\author{Michael Bonitz\textsuperscript{2,3}}
\author{Karsten Balzer\textsuperscript{4}}
\author{Fabio Caruso\textsuperscript{2,3}}
\author{Jan-Philip Joost\textsuperscript{2,3 $\dagger$}}

\dedication{}

\begin{affiliations}
\textsuperscript{1}Max-Planck Institute for Dynamics and Selforanization\\
\textsuperscript{2}Institut f\"ur Theoretische Physik und Astrophysik, Christian-Albrechts-Universit\"at zu Kiel, 24098 Kiel, Germany\\
\textsuperscript{3}Kiel Nano, Surface and Interface Science (KiNSIS), Christian-Albrechts-Universit\"at zu Kiel, Germany\\
\textsuperscript{4}Computing Centre, Christian-Albrechts-Universit\"at zu Kiel, 24098 Kiel, Germany\\
*Email Address: giorgio.lovato@ds.mpg.de\\
$^\dagger$Email Address: joost@theo-physik.uni-kiel.de
\end{affiliations}




\begin{abstract}
The interaction of highly-charged ions with monolayers of graphene and MoS$_2$ is theoretically investigated based on nonequilibrium Green Functions (NEGF). In a recent paper [Niggas et al., Phys. Rev. Lett. \textbf{129}, 086802 (2022)] dramatic differences in the response of the two materials to an impacting slow ion were reported. Here, this analysis is extended, focusing on the effect of electron-electron correlations in the monolayer on the electronic response to the ion. We apply the recently developed time-linear G1--G2 scheme [Schluenzen et al., Phys. Rev. Lett. \textbf{124}, 076601 (2020)] combined with an embedding approach [Balzer et al., Phys. Rev. B \textbf{107}, 155141 (2023)]. We demonstrate that, while electronic correlations have a minor effect in graphene, they significantly influence the electron dynamics in the case of MoS$_2$. Our key results are the ultrafast dynamics of the charge density and induced electrostatic potential in the vicinity of the impact point of the ion.

\end{abstract}


\section{Introduction}\label{s:introduction}
The dynamics of correlated fermions driven out of equilibrium by external forces, such as laser excitation or potential quenches, have attracted high interest in recent years. Examples are the diffusion of fermionic atoms following a confinement quench, e.g. \cite{schneider_np_12,schluenzen_prb16}, and laser excitation of correlated materials in condensed matter \cite{joost_pss_18}. A third example is the response of two-dimensional (2D) quantum materials to the impact of highly-charged ions~\cite{gruber_ultrafast_2016,Niggas_phys-com_21, niggas_prl_22}. This process leads to a very strong excitation of the solid which is very localized in space (on a few lattice sites) and time (a few femtoseconds).
Studying the response of the material to the excitation and the thermalization provides fundamental insights into the electronic and optical properties of the material and may, ultimately, lead to novel technologies in nanoelectronics, such as petahertz switching~\cite{hommelhoff_nat_22}. 

A theoretical description of these processes is challenging because it requires simultaneous account of electronic correlations and quantum effects in the material, energy and charge transfer between the monolayer and the ion, as well as resolution of all processes on sub-femtosecond time scales.
The direct solution of such many-body systems via exact diagonalization is not feasible due to the exponential growth of computational cost with the size of the system. Therefore, approximations have to be developed. Here, the methods of choice and the relevant approximations vary from system to system. In this article we will concentrate on the interaction of highly-charged ions with 2D quantum materials and focus on the methods that have been used for these systems.

A key approximation is Hartree-Fock (HF) which includes many-particle effects via an effective potential (mean field) created by all particles~\cite{hartree_wave_1928,fock_naherungsmethode_1930,slater_simplified_1954}. HF is computationally advantageous and can, in some cases, already capture many of the interaction-based effects in many-body systems. However, correlation effects are neglected which are important, e.g. for two-electron collisions, Auger ionization or interatomic Coulombic decay (ICD) \cite{averbukh_prl_04}. The most 
 popular approach to include correlations is time-dependent density functional theory (TDDFT)~\cite{runge_density-functional_1984}. In this method an exchange-correlation functional of the particle density is introduced in order to include deviations from the Hartree solution. While TDDFT allows to efficiently treat real materials, 
 its capability to describe quasiparticle excitations, correlation effects, and their dynamics is contingent to a suitable choice of the exchange-correlation kernel, for which systematic approximation schemes remain elusive.

 Successful alternative methods are reduced density operators~\cite{bonitz_qkt, kremp-etal.97ap, donsa_prr_23} and non-equilibrium Green function theory (NEGF). Here we focus on NEGF because it provides  systematic approximation schemes, via Feynman diagrams, to correlations and, moreover, direct access to  dynamical and spectral properties of the system~\cite{schluenzen_jpcm_19,stefanucci_many-body_2013,bonitz_qkt}. However, these advantages come with a high computational effort: cubic scaling of the computation time with the number of time steps, $N_t$, as a result of the two time arguments of the one-particle Green function and the memory time integral, limiting the method to short simulations and/or small systems.\\

  These bottlenecks could be overcome to a large extent with the use of the generalized Kadanoff--Baym ansatz (GKBA)~\cite{lipavsky_generalized_1986, bonitz-etal.96jpcm,bonitz_qkt,hermanns_prb14} where the time propagation of the Green functions is restricted to the time diagonal. In particular, the derivation of the time-local G1--G2 scheme~\cite{schluenzen_prl_20,joost_prb_20,joost_prb_22} has led to recent breakthroughs in the computational capabilities, for a recent overview, see Ref.~\cite{bonitz_pssb23}. With this scheme it is now possible to perform quantum many-body simulations of non-uniform systems with sub-femtosecond time resolution with time-linear scaling ($\propto N_t$) \cite{schluenzen_prl_20}.
While the G1--G2 scheme is closely linked to reduced density matrix (RDM) theory derived from the BBGKY-hierarchy~\cite{bogolyubov_problems_1960,boer_studies_1962,joost_prb_22}, it provides direct access to important common NEGF approximation schemes, including the T-matrix or GW-approximations and, therefore, a sound many-body foundation. 

While, with the G1--G2 scheme, the CPU time problem of the propagation of correlated electron systems has been overcome, the remaining main computational bottleneck is the costly storage of the two-particle Green function, $G_2$, for a recent detailed analysis, see Ref.~\cite{bonitz_pssb23}. The main strategy here is either to reduce the basis dimension or to use a basis where the pair interaction is highly diagonal. An example where this is the case is the Hubbard model or its various extensions, such as the Pariser-Parr-Popel (PPP) model. In recent years, Hubbard-type models were successfully applied to many interesting situations of correlated fermion physics, such as confinement quench-excited diffusion of correlated fermionic atoms \cite{schluenzen_prb16,schluenzen_prb17}, laser excitation of graphene~\cite{bonitz_pssb23,joost_phd_2022}, and ion stopping in correlated materials \cite{balzer_prb16,balzer_prl_18,niggas_prl_22}. With a proper selection of the model parameters, usually derived from DFT, high accuracy for the electron dynamics can be achieved~\cite{schluenzen_prb16,schluenzen_prb17}. Therefore, in the present paper we will use a Hubbard model to describe the interaction of monolayers of quantum materials with a highly-charged ion.
A second complementary approach to reduce the size of the interaction matrix and of $G_2$ is to apply an embedding approach \cite{stefanucci_many-body_2013,Bonitz_fcse_19}. The idea is to split the system into two coupled subsystems of which only one (the ``main'' system that is of primary interest) is treated on an advanced many-body level whereas for the rest (the ``environment'') a simpler (and less costly) approximation is being used. In fact, an embedding concept was successfully used to model the neutralization of highly-charged ions \cite{balzer_cpp_21}. Recently the embedding ansatz has been combined with the G1--G2 scheme such that the time-linear scaling  of the full simulation is retained~\cite{kwok_nl_19, balzer_prb_23,bonitz_pssb23,tuovinen_prl_23} which is of high relevance also for transport simulations. \\

Thus, in our present analysis we use the G1--G2 scheme, in combination with a time-linear embedding scheme, in order to verify and extend the theoretical results of Ref.~\cite{niggas_prl_22}. There it was found that graphene and MoS$_2$ show drastic differences in their reaction to the impact of slow highly-charged ions. The most striking one is a 7-fold higher number of secondary electrons that are emitted from a graphene monolayer in comparison to MoS$_2$. The theoretical analysis of Ref.~\cite{niggas_prl_22} that used NEGF simulations on the Hartree-Fock level explained this strong difference in the behavior of the two materials by the much higher electron mobility in graphene. The consequence was that, in graphene, electron loss due to resonant charge transfer to the approaching was compensated more quickly than in MoS$_2$. However, the reliability of this conclusion remains to be tested.

Therefore, the main goal of the present paper is to study the influence of electronic correlations that were neglected in Ref.~\cite{niggas_prl_22}, on the ion-monolayer interaction \cite{lovato_msc_2024}. While correlation effects are known to be moderate in graphene, which has no bandgap, in monolayers of semiconducting MoS$_2$, which have a bandgap of around 1.9\,eV~\cite{splendiani_emerging_2010}, a significant influence is expected. With this property MoS$_2$ can be regarded as a prototype for many 2D quantum materials with high prospects to applications in nanotechnology, e.g.,~\cite{wang_electronics_2012}. So exploring their response to a strong excitation with sub-femtosecond time resolution is of high relevance for other excitation schemes of quantum materials.

 This paper is structured as follows. In Sec.~\ref{s:theoretical_framework}, we derive and discuss the theoretical framework of the G1--G2 scheme, including the time-local embedding scheme. In Sec.~\ref{s:stopping} we apply the derived equations to ion stopping in graphene and MoS$_2$ monolayers using Hubbard-type lattice models  in combination with a model for the resonant charge transfer between the ion and the monolayer. In Sec.~\ref{s:results} we present the numerical results which include the  charge density dynamics in the monolayer during the ion impact. We then compute the electrostatic potential that is produced by this strongly non-uniform charge density in the  the monolayer which has been found to be crucial for the emission of secondary electrons \cite{niggas_prl_22}. Finally, we present results for the spectral properties during the ion impact. 
 The results are summarized in Sec.~\ref{s:discussion} where we also give an outlook for future developments.

 \section{Theoretical Framework}\label{s:theoretical_framework}
 \subsection{Second quantization}
 Our starting point is the nonequilibrium version of  second quantization that is based  on the time evolution of operators in the Heisenberg picture. The key operators to describe a many-body system are the creation ($\hat{c}\U{i}^\dagger$) and annihilation ($\hat{c}\U{i}$) operators, which create or annihilate one particle in orbital $\ket{i}$ of an $N$-particle state. For a given Hamiltonian $\hat{H}$, their dynamics are governed by the Heisenberg equations of motion, namely
 \begin{align*}
     i\hbar\frac{\partial}{\partial t}\hat{c}\U{i}(t)=\big[\hat{H},\hat{c}\U{i}(t)\big],\quad i\hbar \frac{\partial}{\partial t}\hat{c}_i^\dagger(t)=-\big[\hat{c}_i^\dagger(t),\hat{H}\big].
 \end{align*}
 The spin statistics of the system are directly encoded in their (anti-) commutation relations for bosons (fermions)
 \begin{align*}
     \big[\hat{c}_i^\dagger,\hat{c}_j\big]_{\mp}=\hat{c}_i^\dagger\hat{c}_j\mp\hat{c}_j\hat{c}_i^\dagger=\delta_{ij},\quad\big[\hat{c}_i^\dagger,\hat{c}_j^\dagger\big]_{\mp}=\big[\hat{c}_i,\hat{c}_j\big]_{\mp}=0\,,
 \end{align*}
where the upper (lower) signs refer to bosons (fermions). Any quantum mechanical operator can be represented in terms of products of creation and annihilation operators. In the following, we will use the second quantization framework to formulate the dynamics of the system in terms of the most general operators constructed from these products, the non-equilibrium Green functions (NEGF).

  \subsection{Non-equilibrium Green functions}\label{s:non-equilibrium_greens_functions}
In this section, we  give a brief introduction into NEGFs and outline the derivation of the G1--G2 scheme. More information and details on the formalism can be found in Refs.~\cite{joost_prb_20, joost_prb_22,bonitz_pssb23}.\\
 We start with the Hamiltonian of the electron system in second quantization [from now on, we  restrict ourselves to fermions], which can be stated in terms of the creation and annihilation operators, $\hat c_i^\dagger$ and $\hat c_i$, in an arbitrary single-particle basis $\{\ket{i}\}$,
 \begin{align*}
	 \hat{H}(t)=\sum\U{ij}h\Ho{0}\U{ij}(t)\,\hat{c}\U{i}^\dagger\hat{c}\U{j}+\frac{1}{2}\sum\U{ijkl}w\U{ijkl}(t)\hat{c}\U{i}^\dagger\hat{c}\U{j}^\dagger\hat{c}\U{l}\hat{c}\U{k}.
 \end{align*}
 In this expression $h\Ho{0}(t)$ denotes the one-particle Hamiltonian, containing the kinetic energy as well as external potentials, such as the Coulomb potential of the approaching, highly-charged ion~\cite{balzer_prb16,balzer_prl_18}. $w_{ijkl}(t)$ are the matrix elements of the pair interaction potential between electrons. The time dependence reflects the adiabatic switch-on of the potential what allows us to start the simulation from a non-interacting initial state which is known~\cite{schluenzen_jpcm_19}. The definition of the one-particle Green function is given by
 \begin{align*}
	 G\U{ij}(z,z^\prime)=\frac{1}{i\hbar}\big<\mathcal{T}\U{\mathcal{C}}\big\{\hat{c}\U{i}^\dagger(z)\hat{c}\U{j}(z^\prime)\big\}\big>\,,
 \end{align*}
where the arguments $z,z^\prime$ represent time arguments on the Keldysh contour $\mathcal{C}$~\cite{keldysh64,bonitz_pss_19_keldysh}, and $\mathcal{T}\U{\mathcal{C}}$ is the time-ordering operator on the contour. The ensemble averaging indicated by $\braket{\cdot}$ is performed using the correlated unperturbed $N$-particle density operator of the system. 

 The equations of motion for $G$ on the contour  are the Keldysh-Kadanoff-Baym equations (KBEs)~\cite{keldysh64, kadanoff_quantum_2018}
 \begin{align}
	 \sum\U{k}  \bigg[i\hbar\frac{\partial}{\partial z}\delta\U{ik}-h\Ho{0}\U{ik}(z)\bigg]G\U{kj}(z,z^\prime)-\delta\U{ij}\delta\U{\mathcal{C}}(z,z^\prime)              & =-i\hbar\sum\U{klp}\int\U{\mathcal{C}}\mathrm{d}\bar{z}\,w\U{iklp}(z,\bar{z})\,G\Ho{(2)}\U{lpjk}(z,\bar{z},z^\prime,\bar{z}^+)                                                                                                                                                      \label{eq:kbe_g2}    \\
	           & =\sum\U{k}\int\U{\mathcal{C}}\mathrm{d}\bar{z}\,\Sigma\U{ik}(z,\bar{z})\,G\U{kj}(\bar{z},z^\prime),                                                                                                                                                                                                                                                                                                                   \label{eq:kbe_selfe} \\
	 \sum\U{k}  G\U{ik}(z,z^\prime)  \bigg[-i\hbar\overleftarrow{\frac{\partial}{\partial z'}}-h\U{kj}\Ho{0}(z')\bigg]-\delta_{ij}\delta\U{\mathcal{C}}(z,z^\prime)                                                                                                                                                        & =-i\hbar\sum\U{klp}\int\U{\mathcal{C}}\mathrm{d}\bar{z}\,G\Ho{(2)}\U{iklp}(z,\bar{z}^-,z^\prime,\bar{z})\,w\U{lpjk}(\bar{z},z^\prime) \notag               \\
	           & =\sum\U{k}\int\U{\mathcal{C}}\mathrm{d}\bar{z}\,G\U{ik}(z,\bar{z})\,\Sigma\U{kj}(\bar{z},z^\prime).\notag
 \end{align}
The left-hand sides of the equations correspond to the free propagation of the one-particle Green function, whereas the right-hand sides couple the dynamics to the rest of the system and include interaction effects via the pair potential. The right-hand sides are written in two ways. First, Eq.~(\ref{eq:kbe_g2}) involves the two-particle Green function
 \begin{align*}
	 G\U{ijkl}\Ho{(2)}(z\U{1},z\U{2},z\U{3},z\U{4})=\frac{1}{(i\hbar)^2}\big<\mathcal{T}\U{\mathcal{C}}\big\{\hat{c}\U{i}^\dagger(z\U{1})\hat{c}\U{j}^\dagger(z\U{2})\hat{c}\U{l}(z\U{4})\hat{c}\U{k}(z\U{3})\big\}\big>\,,
 \end{align*}
the equation of motion of which couples to the three-particle Green function (Martin-Schwinger-hierarchy~\cite{martin_theory_1959}). The other way [Eq.~(\ref{eq:kbe_selfe})] eliminates the two-particle quantity in favor of the one-particle self-energy $\Sigma$. 
For practical applications we switch to real times and consider the correlation functions
 \begin{align*}
	 G\U{ij}\Ho{<}(t,t^\prime) &  =-\frac{1}{i\hbar}\big<\hat{c}^\dagger\U{j}(t^\prime)\hat{c}\U{i}(t)\big>,
     \\
	 G\U{ij}\Ho{>}(t,t^\prime) &  = \frac{1}{i\hbar}\big<\hat{c}\U{i}(t)\hat{c}^\dagger\U{j}(t^\prime)\big>\,.
 \end{align*}
The value of $G^<$ on the time diagonal, yields all single-particle observables, such as the one-particle reduced density matrix (1-pRDM)
 \begin{align*}
	 F\U{ij}(t)=-i\hbar G\U{ji}\Ho{<}(t,t)\,.
 \end{align*}
In addition, we have access to spectral properties which require the Fourier transform, with respect to the difference time $\tau=t^\prime-t$, of the difference, $G\Ho{>}\U{ij}(t,t^\prime)-G\Ho{<}\U{ij}(t,t^\prime)$~\cite{stefanucci_many-body_2013,bonitz_qkt}. However, there is also a cost to this advantage, namely the computationally expensive contour-time integral on the r.h.s. of the KBEs. The computation of this integral results in a cubic scaling of the CPU time with the propagation duration, making long simulations impossible, except for very small systems.\\
 In the following, we will consider two self-energy approximations. The first is the Hartree-Fock self-energy (mean field), which only has a time-diagonal contribution
 \begin{align}
	 \Sigma\Ho{HF}\U{ij}(t)=-i\hbar\sum\U{kl}w\U{ikjl}\Ho{-}(t)G\U{lk}\Ho{<}(t,t)\,,
     \label{eq:sigma-hf}
 \end{align}
 where $w\U{ikjl}\Ho{-} := w\U{ikjl}-w\U{iklj}$.
 Due to the simple structure of the HF self-energy, it can be treated as an effective one-particle Hamiltonian and grouped together with the single-particle Hamiltonian
 \begin{align*}
	 h\U{ij}\Ho{HF}(t)=h\U{ij}\Ho{0}-i\hbar\sum\U{kl}w\U{ikjl}\Ho{-}(t)G\U{lk}\Ho{<}(t,t).
 \end{align*}
 The simplest self-energy beyond Hartree-Fock that includes correlation effects is the second-order Born self-energy which does include time-off-diagonal elements,
 \begin{align}
	 \Sigma\Ho{\gtrless,SOA}\U{ij}(t,t^\prime)=-(i\hbar)^2\sum\U{klpqrs}w\U{iklp}\Ho{-}(t)w\U{qrjs}\Ho{-}(t^\prime)G\U{lq}\Ho{\gtrless}(t,t^\prime)G\U{pr}\Ho{\gtrless}(t,t^\prime)G\U{sk}\Ho{\lessgtr}(t^\prime,t).
     \label{eq:sigma-soa}
 \end{align}
 When speaking about the second-order Born approximation (SOA) below, this  will include the correlation self-energy (\ref{eq:sigma-soa})  
 as well as the HF-contribution (\ref{eq:sigma-hf}).
 
 \subsection{HF-GKBA and G1--G2 scheme}\label{s:hf-gkba}
 In some applications, it is not unreasonable to discard accurate spectral information in exchange for a less expensive propagation. In fact, we can even go one step further by considering only the evolution of the one-particle Green function along the time diagonal
 \begin{align}
	 i\hbar\dt G\U{ij}\Ho{<}(t)-\big[\bs{h}\Ho{HF},\bs{G}\Ho{<}\big]\U{ij}(t)=i\hbar\sum\U{k}\int\U{t_0}\Ho{t}\mathrm{d}\,\bar{t}\,\left[\Sigma\Ho{SOA,>}\U{ik}(t,\bar{t}\,)G\U{kj}\Ho{<}(\bar{t},t)-\Sigma\Ho{SOA,<}\U{ik}(t,\bar{t}\,)G\U{jj}\Ho{>}(\bar{t},t)\right]\,,
     \label{eq:kbe-diagonal}
 \end{align}
 where we denoted matrices in orbital space by bold letters and used the short notation $[\textbf{A},\textbf{B}]_{ij}:=\sum_k(A_{ik}B_{kj}-B_{ik}A_{kj})$.
 Since the evolution on the time diagonal still depends on the off-diagonal elements, under the time integral, the equation of motion (EOM) is not fully time-diagonal and the costly memory integral still occurs.\\
 For this reason, a common way to approximate the full two-time solution is to use a reconstruction of the off-diagonal part of the one-particle Green function~\cite{lipavsky_generalized_1986}
 \begin{align}\label{eq:g1_reconstruction}
	 G\U{ij}^{\gtrless}(t,t^\prime)=i\hbar\sum\U{k}\big\{G\Ho{R}\U{ik}(t,t^\prime)G^\gtrless\U{kj}(t^\prime)-G^\gtrless\U{ik}(t)G\Ho{A}\U{kj}(t,t^\prime)\big\},
 \end{align}
 with the retarded and advanced Green functions being defined as
 \begin{align*}
	 G\Ho{R/A}\U{ij}(t,t^\prime)=\pm\Theta[\pm(t-t^\prime)]\big\{G\U{ij}\Ho{>}(t,t^\prime)-G\U{ij}\Ho{<}(t,t^\prime)\big\}.
 \end{align*}
 The reconstruction already represents an approximation since additional terms, involving time integrals as well as the self-energy, are neglected \cite{lipavsky_generalized_1986}. The ansatz (\ref{eq:g1_reconstruction}) for the off-diagonal parts is commonly known as the Generalized Kadanoff--Baym Ansatz (GKBA). To apply this reconstruction, we also need to know the time-off-diagonal elements of the propagators $G\Ho{R/A}(t,t')$. The most common choices are either the free GKBA or the HF-GKBA, where the propagators are the free and the Hartree-Fock propagators, respectively. In SOA, this approximation reduces the computational cost significantly (to quadratic in the number of time steps), but for higher-order self-energies the cubic scaling remains. Furthermore, due to the use of HF propagators,  spectral properties do not contain correlation effects.
 
 As described in~Ref.~\cite{schluenzen_prl_20} and further expanded upon in Refs.~\cite{bonitz_pssb23,joost_prb_20,joost_prb_22}, one can, in fact, by using the GKBA, completely remove the memory integral from the EOM for the Green function on the time diagonal. This is achieved by utilizing the KBEs with collision integrals defined in terms of the two-particle Green function, Eq.~(\ref{eq:kbe_g2}), instead of the self-energy formulation of Eq.~(\ref{eq:kbe_selfe}). The strategy is fairly straightforward:
 \begin{itemize}
	 \item[I.)] Choose a specific self-energy and insert it into the KBEs on the time-diagonal.
	 \item[II.)] Compare the expressions on the right-hand side to the first equation of the Martin-Schwinger-hierarchy\footnote{Meaning the KBEs without the self-energy.}.
	 \item[III.)] Identify the appropriate formal solution of $G\Ho{(2)}$ in terms of $G^<$.
	 \item[IV.)] Using the HF-GKBA, derive the EOM for $G\Ho{(2)}$.
 \end{itemize}
 The result is the G1--G2 scheme which has been proven to exhibit linear scaling in the propagation time at the cost of propagating, in addition to $G^<$, the two-particle Green function~\cite{schluenzen_prl_20}.\\
 To illustrate this procedure, we recall the main steps of the derivation of the G1--G2 scheme in SOA. The EOM for $G^<$ on the diagonal reads
 \begin{align}
	 i\hbar\frac{\partial}{\partial t} G\Ho{<}\U{ij}(t)-\big[\bs{h}\Ho{0},\bs{G}\Ho{<}\big]\U{ij}(t) & =\sum\U{k}\Sigma\U{ik}\Ho{HF}(t)G\Ho{<}\U{kj}(t)\notag \\
	 & \quad-\sum\U{k}\int\U{t\U{t}}\Ho{t}\mathrm{d}\bar{t}\left[\Sigma\Ho{>,SOA}\U{ik}(t,\bar{t})G\Ho{<}\U{kj}(\bar{t},t)-\Sigma\U{ik}\Ho{<,SOA}(t,\bar{t})G\Ho{>}\U{kj}(\bar{t},t)\right] \label{eq:kbe_rt_selfe} \\
    & =-i\hbar\sum\U{klp}w\U{iklp}(t)G\Ho{(2)}\U{lpjk}(t).\label{eq:kbe_rt_g2}
 \end{align}
 Next, we split $G\Ho{(2)}$ into two parts,
 \begin{align*}
	 G\Ho{(2)}\U{ijkl}(t)=G\Ho{(2),HF}\U{ijkl}(t)+\mathcal{G}\U{ijkl}(t)\,,
 \end{align*}
 where the first and second parts account for the HF and SOA self-energies, respectively. By comparing Eqs.~(\ref{eq:kbe_rt_selfe}) and (\ref{eq:kbe_rt_g2}), we can conclude the following identities
 \begin{align}
	 G\U{ijkl}\Ho{\gtrless,(2),HF}(t) =G\U{ik}\Ho{\gtrless}(t)G\U{jl}\Ho{\gtrless}(t)-G\U{il}\Ho{\gtrless}(t)G\U{jk}\Ho{\lessgtr}(t)=G\U{ijkl}\Ho{\gtrless,(2),H}(t)-G\U{ijkl}\Ho{\gtrless,(2),F}(t)\,,
     \label{eq:g2h-g2f}
 \end{align}
 and
 \begin{align*}
	 \mathcal{G}\U{ijkl}(t) & =i\hbar\sum\U{pqrs}\int\U{t\U{0}}\Ho{t}\mathrm{d}\bar{t}w\U{pqrs}\Ho{-}(\bar{t}\,)\big[G\U{ijpq}\Ho{>,(2),H}(t,\bar{t})G\U{rskl}\Ho{>,(2),H}(\bar{t},t)-G\U{ijpq}\Ho{<,(2),H}(t,\bar{t})G\U{rskl}\Ho{<,(2),H}(\bar{t},t)\big]\,,
 \end{align*}
 where, in Eq.~\eqref{eq:g2h-g2f}, we introduced the two-particle Hartree and Fock Green functions.
 Finally, by applying the HF-GKBA and differentiating the expression of $G\Ho{(2)}$ with respect to time we obtain the coupled set of EOMs for the G1--G2 scheme in SOA~\cite{joost_phd_2022}
 \begin{align*}
	 i\hbar\DT G\U{ij}\Ho{<}(t)-\big[\bs{h}\Ho{HF},\bs{G}\Ho{<}\big]\U{ij}(t)                & =\big[\bs{I}+\bs{I}\Ho{\dagger}\big]\U{ij}(t),\quad I\U{ij}(t)=-i\hbar\sum\U{klp}w\U{iklp}(t)\mathcal{G}\U{lpjk}(t),                                          \\
	 i\hbar\DT\mathcal{G}\U{ijkl}(t)-\big[\bs{h}\Ho{(2),HF},\bs{\mathcal{G}}\big]\U{ijkl}(t) & =\frac{1}{2}\sum\U{pq}w\U{ijpq}\Ho{-}(t)G\U{pk}\Ho{<}(t)G\U{ql}\Ho{<}(t)+i\hbar\sum\U{pqr}w\U{ipqr}\Ho{-}(t)G\U{jp}\Ho{<}(t)G\U{qk}\Ho{<}(t)G\U{rl}\Ho{<}(t).
 \end{align*}
 Using the same strategy, one can find EOMs for all higher-order self-energies, as demonstrated in~\cite{joost_prb_22}. There, comparisons between the RDM and the G1--G2 approach were performed. The authors conclude that both methods are equivalent in HF, SOA and TPP approximation with and without exchange, meaning that for all known NEGF self-energies one can find an equivalent G1--G2 formulation. Since approximations stemming from self-energies can be better understood in terms of the diagrams they include, it is very useful to be able to assign them to the time-diagonal contributions. At the same time, the G1--G2 scheme allows for additional many-body approximations for which no direct single-particle self-energies are known. An example is the dynamically screened ladder (DSL) approximation, for details, see Ref.~\cite{joost_prb_22}. 

 In this work we will focus on the HF and SOA self-energies and make use of the G1--G2 framework which has been tested multiple times for simple Hubbard chains and more complex finite graphene clusters~\cite{joost_prb_20,joost_prb_22}.

 \subsection{NEGF embedding approach}\label{s:embedding}
 In the following, we focus on a special case of the G1--G2 scheme, namely the description of a total system consisting of two distinct parts which will be called ``system'' and ``environment''. Possible applications for such a framework are the description of electron dynamics in a crystal with metal contacts or, as we will discuss later in Sec.~\ref{s:stopping},  an ion colliding with a graphene or MoS$_2$ monolayer.
 Of course the G1--G2 scheme could be used to describe the full system in detail but that might require unnecessary computation cost since the system and the environment might evolve on completely different time scales and correlation effects might not be of equal importance in the two parts. The goal of the embedding scheme is, therefore, to be able to employ different approximations for the correlations in both, the system and the environment. For example, in the ion stopping scenario, one might want to use a higher-order self-energy, such as SOA, in the system (the target),  while treating the ion without correlations altogether.
 The derivation of such a framework from the NEGF perspective is demonstrated, e.g., in Refs.~\cite{stefanucci_many-body_2013, balzer_cpp_21,balzer_prb_23}. 
 
 \subsubsection{Embedding Hamiltonian}
 The Hamiltonian of the total system can be written in the following form \cite{Bonitz_fcse_19}
 \begin{align*}
	 \hat{H}(t)=\sum\U{\alpha,\beta\in\Omega}\sum\U{ij}h\U{ij}\Ho{\alpha\beta}(t)\hat{c}\U{i}\Ho{\alpha,\dagger}\hat{c}\U{j}\Ho{\beta,\dagger}+\frac{1}{2}\sum\U{\alpha,\beta,\gamma,\delta\in\Omega}\sum\U{ijkl}w\U{ijkl}\Ho{\alpha\beta\gamma\delta}(t)\hat{c}\U{i}\Ho{\alpha,\dagger}\hat{c}\U{j}\Ho{\beta,\dagger}\hat{c}\U{l}\Ho{\delta}\hat{c}\U{k}\Ho{\gamma}\,,
 \end{align*}
 where $\Omega=\{\textup{e},\textup{s}\}$ is a set of indices with $e$ denoting states of the environment and $s$ states of the system (this assumes that a physical separation is possible). The creation and annihilation operators likewise refer to state $i$ of system part $\alpha$ and create or annihilate a particle in this specific state. The one-particle Hamiltonian now includes three different parts
 \begin{align*}
	 \bs{h}=
	 \begin{pmatrix}
		 \bs{h}\Ho{ss} & \bs{h}\Ho{se} \\
		 \bs{h}\Ho{es} & \bs{h}\Ho{ee}
	 \end{pmatrix},
 \end{align*}
 (with $h\Ho{se}=h\Ho{es,\dagger}$): the one-particle Hamiltonian of the system ($ss$), of the environment ($ee$) and the coupling between the two ($es$). The same applies to the interaction pair potential $w$, the self-energy, and the one-particle Green function
 \begin{align*}
	 G\U{ij}\Ho{\alpha\beta}(z,z^\prime)=\frac{1}{i\hbar}\big<\mathcal{T}\U{\mathcal{C}}\big\{\hat{c}\Ho{\alpha,\dagger}\U{i}(z)\hat{c}\U{j}\Ho{\beta}(z^\prime)\big\}\big>.
 \end{align*}
 This approach will lead to a set of four KBEs, for all combinations of $\alpha$ and $\beta$,
 \begin{align*}
	 i\hbar\dz G\U{ij}\Ho{\alpha\beta}(z,z^\prime)-\sum\U{\delta\in\Omega}\sum\U{k}h\U{ik}\Ho{\alpha\delta,(HF)}(z)G\U{kj}\Ho{\delta\beta}(z,z^\prime)=\delta\U{ij}\Ho{\alpha\beta}\delta\U{\mathcal{C}}(z,z^\prime)+\sum\U{\delta\in\Omega}\sum\U{k}\int\U{\mathcal{C}}\mathrm{d}\bar{z}\,\Sigma\Ho{\alpha\delta}\U{ik}(z,\bar{z})G\U{kj}\Ho{\delta\beta}(\bar{z},z^\prime)\,,
 \end{align*}
 which is the starting point for approximations and reformulations.
 
 \subsubsection{NEGF embedding equations}\label{s:NEGF_embedding_equations}
Following the discussion above, we regard the 
``environment'' as part of the total system inside of which correlation effects are of minor importance. Therefore, in the following, we neglect the correlation contributions of the self-energy within in the environment, as well as in the transfer between system and environment. Assuming again separability of the system in two parts, the simplest choice is
 \begin{align*}
	 \bs{\Sigma}\Ho{es}=\bs{\Sigma}\Ho{ee}=0,\quad \bs{\Sigma}\Ho{ss}=\bs{\Sigma}\,,
 \end{align*}
 where we drop the superscripts, in the following.
 This ansatz enables us to write the equations of motion for the whole system as~\cite{balzer_prb_23}
 \begin{align}
	 i\hbar\dz G\U{ij}\Ho{ss}(z,z^\prime)-\sum\U{k}h\U{ik}\Ho{ss,(HF)}(z)G\U{kj}\Ho{ss}(z,z^\prime) & =\delta\U{ij}\delta\U{\mathcal{C}}(z,z^\prime)+\sum\U{k}\int\U{\mathcal{C}}\mathrm{d}\bar{z}\big\{\Sigma\U{ik}(z,\bar{z})+\Sigma\U{ik}\Ho{emb}(z,\bar{z})\big\}G\U{kj}\Ho{ss}(\bar{z},z^\prime),
     \label{eq:gs-equation-emb}\\
	 \Sigma\Ho{emb}\U{ij}(z,z^\prime)                                                               & =\sum\U{kl}h\U{ik}\Ho{se,(HF)}(z)G\U{kl}\Ho{ee}(z,z^\prime)h\U{lj}\Ho{es,(HF)}(z^\prime), \label{eq:sigma-emb-2time}\\\nonumber
	 i\hbar\dz G\U{ij}\Ho{ee}(z,z^\prime)-\sum\U{k}h\U{ik}\Ho{ee,(HF)}(z)G\U{kj}\Ho{ee}(z,z^\prime) & =\sum\U{k}h\U{ik}\Ho{es,(HF)}(z)G\U{kj}\Ho{se}(z,z^\prime)+\delta\U{ij}\delta\U{\mathcal{C}}(z,z^\prime),
    \\\nonumber
	 i\hbar\dz G\U{ij}\Ho{es}(z,z^\prime)& =\sum\U{k}h\U{ik}\Ho{es,(HF)}(z)G\U{kj}\Ho{ss}(z,z^\prime)-\sum\U{k}h\U{ik}\Ho{ee,(HF)}(z)G\U{kj}\Ho{es}(z,z^\prime)\, .
 \end{align}
The embedding approach allows one to derive a closed equation for the system Green function, Eq.~\eqref{eq:gs-equation-emb}, where the e-s coupling is exactly accounted for via an additional ``embedding'' self-energy, Eq.~(\ref{eq:sigma-emb-2time}), see also Refs.~\cite{myohanen_prb_09,stefanucci_many-body_2013}. However, this equation still has the unfavorable cubic scaling of the CPU-time with $N_t$ that is inherent in the two-time KBE.
 
Therefore, it is again advantageous to derive a time-local G1--G2 version of the embedding scheme, as was demonstrated in Ref.~\cite{balzer_prb_23}. The main steps are again:  transition to the real-time plane, application of the HF-GKBA and differentiation of the resulting expressions with respect to time. The result is the following coupled set of time-local EOMs
 \begin{align}
	  & i\hbar\dt G\U{ij}\Ho{ss,<}(t)-\big[\bs{h}\Ho{ss,(HF)},\bs{G}\Ho{ss,<}\big]\U{ij}(t)=\big[\bs{I}(t)+\bs{I}^\dagger(t)\big], \label{eq:eom_g1g2_emb_ss}\\
	  & \bs{I}(t)=\bs{I}\Ho{corr}(t)+\bs{I}\Ho{emb}(t),\quad I\Ho{corr}\U{ij}(t)  =-i\sum\U{klp}w\U{iklp}(t)\mathcal{G}\U{lpjk}(t),\quad I\Ho{emb}\U{ij}(t)=\sum\U{k}h\U{ik}\Ho{se,(HF)}(t)G\U{kj}\Ho{es,<}(t),\notag \\
	  & i\hbar\dt G\U{ij}\Ho{es,<}(t)=\sum\U{k}h\U{ik}\Ho{es,(HF)}(t)G\U{kj}\Ho{ss,<}(t)-\sum\U{k}G\U{ik}\Ho{ee,<}(t)h\U{kj}\Ho{es,(HF)}(t)\notag\\
	  & \quad\quad\quad\quad\quad\quad+\sum\U{k}h\U{ik}\Ho{ee,(HF)}(t)G\U{kj}\Ho{es,<}(t)-\sum\U{k}G\U{ik}\Ho{es,<}(t)h\U{kj}\Ho{ss,(HF)}(t), \label{eq:eom_g1g2_emb_es}\\
	  & i\hbar\dt G\U{ij}\Ho{ee,<}(t)-\big[\bs{h}\Ho{ee,(HF)},\bs{G}\Ho{ee,<}\big]\U{ij}(t)=\sum\U{k}h\U{ik}\Ho{es,(HF)}(t)G\U{kj}\Ho{se,<}(t)-\sum\U{k}G\U{ik}\Ho{es,<}(t)h\U{kj}\Ho{se,(HF)}(t),\label{eq:eom_g1g2_emb_ee}\\
	  & i\hbar\dt \mathcal{G}\U{ijkl}(t)-\big[\bs{h}\Ho{(2),(HF)},\bs{\mathcal{G}}\big]\U{ijkl}(t)=C\U{ijkl}(t).\notag
 \end{align}
 Here, we used the shorthand $C\U{ijkl}(t)$ to denote any kind of approximation for the two-particle interactions one might choose in the system. Note also that we do not have to define an EOM for $G\Ho{se,<}$ since
 \begin{align*}
	 \bs{G}\Ho{se,<}(t)=-\big[\bs{G}\Ho{es,<}\big]^\dagger(t).
 \end{align*}
 For details of the derivation and conservation properties, we refer to Ref.~\cite{balzer_prb_23}; a similar result has been obtained in Ref.~\cite{tuovinen_prl_23}.
 
 \subsection{Derivation of the embedding equations in a time-local framework}\label{s:time-local_embedding}
 An alternative route to the time-linear G1--G2 embedding scheme can be taken if one starts directly with the conventional EOM for the real-time single-particle Green function on the time diagonal,
 \begin{align}
	 i\hbar\dt G\U{ij}\Ho{<}(t)-\big[\bs{h}\Ho{(HF)},\bs{G}\Ho{<}\big]\U{ij}(t)=\big[\bs{I}(t)+\bs{I}^\dagger(t)\big]\U{ij},\label{eq:eom_g1g2_1}
 \end{align}
 where the collision integral $I$ contains the coupling to the correlation part, $\mathcal{G}$, of the two-particle Green function, which we will put aside for the moment. Consider now a reinterpretation of the basis indices $i,j$ as referring to states that either live in the system or the environment
 \begin{align*}
	 \ket{i}=\ket{\alpha,\Ul{i}},\quad \alpha\in\Omega=\{\textup{e},\textup{s}\},\Ul{i}\in\{\Ul{i}\,,\ket{\Ul{i}\,}\in \textup{e}\}.
 \end{align*}
 In this representation the one-particle Green function and all other single-particle quantities become $2\times 2$-matrices,
 \begin{align*}
	 \bs{G}\Ho{<}\rightarrow
	 \begin{pmatrix}
		 \bs{G}\Ho{ss,<} & \bs{G}\Ho{se,<} \\
		 \bs{G}\Ho{es,<} & \bs{G}\Ho{ee,<}
	 \end{pmatrix},\quad
	 \bs{h}\Ho{(HF)}\rightarrow
	 \begin{pmatrix}
		 \bs{h}\Ho{ss,(HF)} & \bs{h}\Ho{se,(HF)} \\
		 \bs{h}\Ho{es,(HF)} & \bs{h}\Ho{ee,(HF)}
	 \end{pmatrix}\,\,\mathrm{and}\,\,
	 \bs{I}\rightarrow
	 \begin{pmatrix}
		 \bs{I}\Ho{ss} & \bs{I}\Ho{se} \\
		 \bs{I}\Ho{es} & \bs{I}\Ho{ee}
	 \end{pmatrix}.
 \end{align*}
 Inserting this into Eq.~(\ref{eq:eom_g1g2_1}) we obtain
 \begin{align*}
   i\hbar\dt G\U{\Ul{i}\Ul{j}}\Ho{\alpha\beta,<}(t)-\sum\U{\delta\in\Omega}\sum\U{\Ul{k}}\big(h\U{\Ul{i}\Ul{k}}\Ho{\alpha\delta,(HF)}(t)G\U{\Ul{k}\Ul{j}}\Ho{\delta\beta,<}(t)-G\U{\Ul{j}\Ul{k}}\Ho{\alpha\delta,<}(t)h\U{\Ul{k}\Ul{j}}\Ho{\delta\beta,(HF)}(t)\big)=\big[\bs{I}(t)+\bs{I}^\dagger(t)\big]\U{\Ul{i}\Ul{j}}\Ho{\alpha\beta}.
 \end{align*}
 Now we can simply read off the EOM for the ($\textup{ss}$) component of the single-particle Green function
 \begin{align*}
	  & i\hbar\dt G\U{\Ul{i}\Ul{j}}\Ho{ss,<}(t)-\sum\U{\Ul{k}}\big(h\U{\Ul{i}\Ul{k}}\Ho{ss,(HF)}(t)G\U{\Ul{k}\Ul{j}}\Ho{ss,<}(t)-G\U{\Ul{i}\Ul{k}}\Ho{ss,<}(t)h\U{\Ul{k}\Ul{j}}\Ho{ss,(HF)}(t)\\
	  & \quad\quad\quad\quad\quad\quad\quad\quad+h\U{\Ul{i}\Ul{k}}\Ho{se,(HF)}(t)G\U{\Ul{k}\Ul{j}}\Ho{es,<}(t)-G\U{\Ul{i}\Ul{k}}\Ho{se,<}(t)h\U{\Ul{k}\Ul{j}}\Ho{es,(HF)}(t)\big)\\
	  & \quad=i\hbar\dt G\U{\Ul{i}\Ul{j}}\Ho{ss,<}(t)-\big[\bs{h}\Ho{ss,(HF)},\bs{G}\Ho{ss,<}\big]\U{\Ul{i}\Ul{j}}(t)-\underbrace{\sum\U{\Ul{k}}h\U{\Ul{i}\Ul{k}}\Ho{se,(HF)}(t)G\U{\Ul{k}\Ul{j}}\Ho{es,<}(t)}_{\bs{I}\Ho{ss,emb}(t)}+\underbrace{\sum\U{\Ul{k}}G\U{\Ul{i}\Ul{k}}\Ho{se,<}(t)h\U{\Ul{k}\Ul{j}}\Ho{es,(HF)}(t)}_{\bs{I}\Ho{ss,emb,\dagger}(t)},
 \end{align*}
 Thus, in total, we obtain
 \begin{align*}
     & i\hbar\dt G\U{\Ul{i}\Ul{j}}\Ho{ss,<}-\big[\bs{h}\Ho{ss,(HF)},\bs{G}\Ho{ss,<}\big]\U{\Ul{i}\Ul{j}}(t)=\big[\bs{I}(t)+\bs{I}^\dagger(t)\big]\U{\Ul{i}\Ul{j}}\Ho{ss}+\big[\bs{I}\Ho{emb}(t)+\bs{I}\Ho{emb,\dagger}(t)\big]\U{\Ul{i}\Ul{j}}\Ho{ss},
 \end{align*}
 which is equivalent to the EOM derived from the NEGF approach, i.e., Eq.~(\ref{eq:eom_g1g2_emb_ss}). The remaining components can be inferred analogously
 \begin{align*}
	 i\hbar\dt G\U{\Ul{i}\Ul{j}}\Ho{ee,<}(t)-\big[\bs{h}\Ho{ee,(HF)},\bs{G}\Ho{ee,<}\big]\U{\Ul{i}\Ul{j}}(t) & =\big[\bs{I}(t)+\bs{I}^\dagger(t)\big]\U{\Ul{i}\Ul{j}}\Ho{ee}\\
     &+\sum\U{\Ul{k}}\big(h\U{\Ul{i}\Ul{k}}\Ho{es,(HF)}(t)G\U{\Ul{k}\Ul{j}}\Ho{se,<}(t)-G\U{\Ul{i}\Ul{k}}\Ho{es,(HF)}(t)h\U{\Ul{k}\Ul{j}}\Ho{se,(HF)}(t)\big)\\
	  & \eqqcolon \big[\bs{I}(t)+\bs{I}^\dagger(t)\big]\U{\Ul{i}\Ul{j}}\Ho{ee}+\big[\bs{I}\Ho{emb}(t)+\bs{I}\Ho{emb,\dagger}(t)\big]\U{\Ul{i}\Ul{j}}\Ho{ee},\\
	 i\hbar\dt G\U{\Ul{i}\Ul{j}}\Ho{es,<}(t)& =\big[\bs{I}(t)+\bs{I}^\dagger(t)\big]\U{\Ul{i}\Ul{j}}\Ho{es}+\sum\U{\Ul{k}}\big(h\U{\Ul{i}\Ul{k}}\Ho{es,(HF)}(t)G\U{\Ul{k}\Ul{j}}\Ho{ss,<}(t)+h\U{\Ul{i}\Ul{k}}\Ho{ee,(HF)}(t)G\U{\Ul{k}\Ul{j}}\Ho{es,<}(t)\big) \\
	 & \quad   -\sum\U{\Ul{k}}\big(G\U{\Ul{i}\Ul{k}}\Ho{es,<}(t)h\U{\Ul{k}\Ul{j}}\Ho{ss,(HF)}(t)+G\U{\Ul{i}\Ul{k}}\Ho{ee,(HF)}(t)h\U{\Ul{k}\Ul{j}}\Ho{es,(HF)}(t)\big)\\
	& \eqqcolon\big[\bs{I}(t)+\bs{I}^\dagger(t)\big]\U{\Ul{i}\Ul{j}}\Ho{es}+\big[\bs{I}\Ho{emb}(t)+\bs{I}\Ho{emb,\dagger}(t)\big]\U{\Ul{i}\Ul{j}}\Ho{es}.
 \end{align*}
 By comparing these equations to Eqs.~(\ref{eq:eom_g1g2_emb_es}) and (\ref{eq:eom_g1g2_emb_ee}), it is clear that they are almost equivalent. The only difference is the appearance of the collision integrals $I\Ho{ee}$ and $I\Ho{es}$, which were previously removed by choosing a specific ansatz for the self-energies in these subspaces. So in this case, to achieve the same results as in the previous section, we neglect all collision integrals except for $I\Ho{ss}$, which takes the usual form
 \begin{align*}
	 I\Ho{ss}\U{ij}(t)=(i\hbar)^2\sum\U{klp}w\U{iklp}(t)\mathcal{G}\U{lpjk}(t),
 \end{align*}
 with the equation of motion for the two-particle function
 \begin{align*}
	 i\hbar\dt\mathcal{G}\U{ijkl}-\big[\bs{h}\Ho{(2),(HF)},\bs{\mathcal{G}}\big]\U{ijkl}(t)=C\U{ijkl}(t).
 \end{align*}
 It is apparent that, in the general case, this is a quite severe approximation, as we did not only neglect the collision integrals but also the coupling between the subspaces in the EOM for $\mathcal{G}$. The treatment is analogous to the non-embedding case and therefore represents the approximation of only including correlational effects in the central system. However, the magnitude of the deviation is not quantified in this approach and should therefore be tested for each specific case. A more thorough study of the different collision integrals and the complete EOMs that include all the $\mathcal{G}\Ho{ssss}$, $\mathcal{G}\Ho{eses}$ and so on, will be performed in future work. 
 
In the following, we use the above equations in the G1--G2 embedding framework  and apply them to the  ion stopping scenario that was discussed before. We will concentrate on ultrafast charge transfer processes that proceed within a few femtoseconds. On these time scales the lattice response is only of minor importance (the phonon period of MoS$_2$ is on the order of 80 fs) and will not be considered here. Phonon processes can be directly incorporated in our time linear scheme as was demonstrated e.g. in Refs.~\cite{stefanucci_prx_23,emeis_prx_25}.

 \section{Ion stopping with charge transfer}\label{s:stopping}

 \subsection{Model Hamiltonians for graphene and MoS$_2$ monolayers}\label{ss:monolayer}
 In Sec.~\ref{s:theoretical_framework}, we introduced the general form of the many-body Hamiltonian. To implement a specific model, we now have to find expressions for the single-particle Hamiltonian, as well as the pair potential. Ultimately we want to describe the electron dynamics in 2D graphene and MoS$_2$ monolayers. For graphene we use the approach that was introduced in Refs.~\cite{balzer_cpp_21,niggas_prl_22}, 
 where the delocalised $\pi$ electrons are treated  by a single-band fourfold degenerate Hubbard model  
 \begin{align*}
     \hat{H}^{\mathrm{ss}}=\sum_{i,\sigma}\big(\epsilon-W_{ii}\big)\hat{c}_{i,\sigma}^\dagger\hat{c}_{i,\sigma}-J\sum_{\braket{i,j},\sigma}\hat{c}_{i,\sigma}^\dagger\hat{c}_{j,\sigma}+U\sum_i\hat{n}_{i,\uparrow}\hat{n}_{i,\downarrow}\,,
 \end{align*}
 where also an external potential $W_{ii}$ is included.
 This model depends solely on three physical parameters: the on-site energy $\epsilon$, the nearest-neighbor hopping amplitude $J$ and the Hubbard interaction strength $U$. The geometry of the lattice is captured by the nearest-neighbor sum, indicated by $\braket{i,j}$ and the full, pairwise Coulomb interaction is approximated by the local, on-site, repulsive energy contribution proportional to the spin densities $\hat{n}_{i,\sigma}$. In this work we use the numerical values $\epsilon=1.22J_0$, $J=0.74J_0$, and $U=1.19J_0$~\cite{niggas_prl_22}, which are obtained by fitting the resulting band structure to DFT results where we adopt the unit system $J_0=\hbar/m_{\mathrm{e}}/a_0^2\approx3.78\mathrm{eV}$, $a_0\approx1.42${\AA}~\cite{balzer_cpp_21}.

 Now we discuss the case of MoS$_2$ in more detail.
While this single-band approach is suitable for single-layer graphene (SLG), MoS$_2$ has a more complex band structure. Efforts to reduce the full band structure into compressed 5-band~\cite{shanavas_effective_2015} and 3-band~\cite{liu_three-band_2013} models have proven successful in reproducing the DFT band structure. However, these models are tight-binding models, i.e. they completely neglect correlations. The over-fitted band structure does not leave room for a simple interaction model, which is necessary to study the effects of correlations in the G1--G2 scheme.
  
When projected onto the plane containing the monolayer, the S-Mo-S unit cells form a 2D honeycomb lattice.
We, therefore, employ the same strategy as in Ref.~\cite{niggas_prl_22} and use the same effective, single-band Hamiltonian to describe four identical, non-interacting electron bands in MoS$_2$. The parameters which provide a good fit for the physical system are $\epsilon=1.19J_0$, $J=0.29J_0$ and $U=1.19J_0$. The lower ratio of $J/U$ in MoS$_2$, compared to graphene, reflects the low carrier mobility in the semiconductor.
 \begin{figure}[h]
     \centering
     \includegraphics[width=0.7\textwidth]{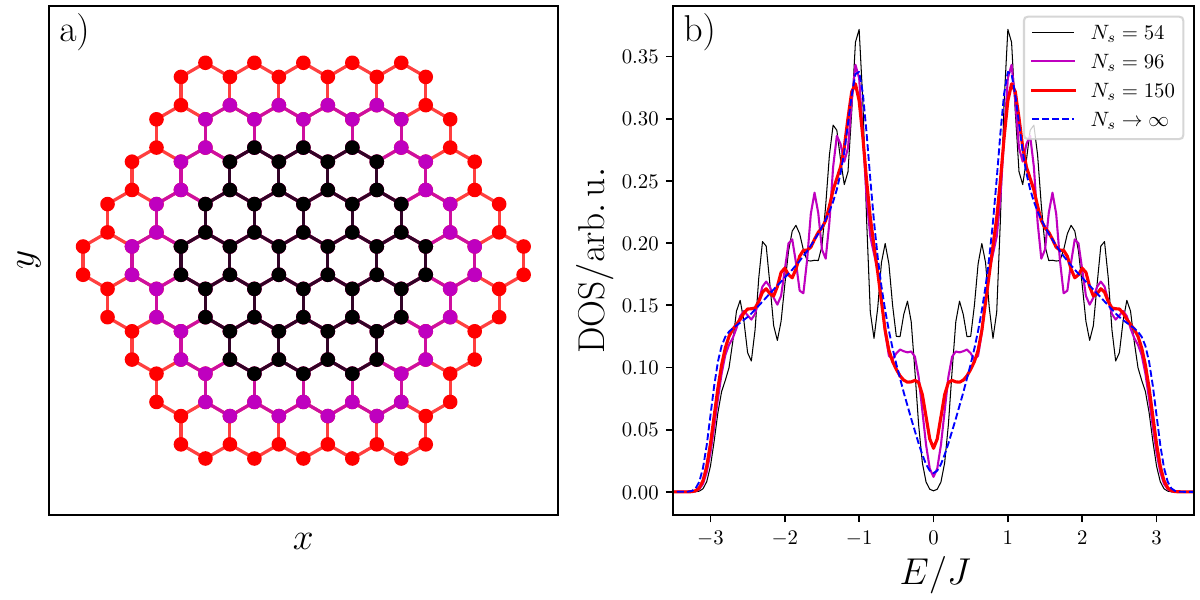}
     \caption{a) Finite honeycomb clusters with a number of $N_s$ sites, b) corresponding DOS compared to the DOS of the infinite lattice (tight binding).}
     \label{fig:dos_explained_3.pdf}
 \end{figure}
 In the absence of interactions ($U=0$) the density of states (DOS) of the system can be obtained directly by diagonalization of the Hamiltonian. An example of honeycomb flakes of different size (colors) are shown in  Fig.~\ref{fig:dos_explained_3.pdf} a). Finite size effects lead to so-called edgestates, which appear as secondary peaks in the DOS (Fig.~\ref{fig:dos_explained_3.pdf}, b)). Their relative magnitude decreases with the system size and the shape of the DOS approaches the TB-DOS for the infinite lattice, also depicted in Fig.~\ref{fig:dos_explained_3.pdf}, b) for comparison.


 \subsection{Model Hamiltonian for the ion stopping and charge transfer}\label{ss:charge-transfer}
 \begin{figure}[h!]
    \centering
    \includegraphics[width=0.7\textwidth]{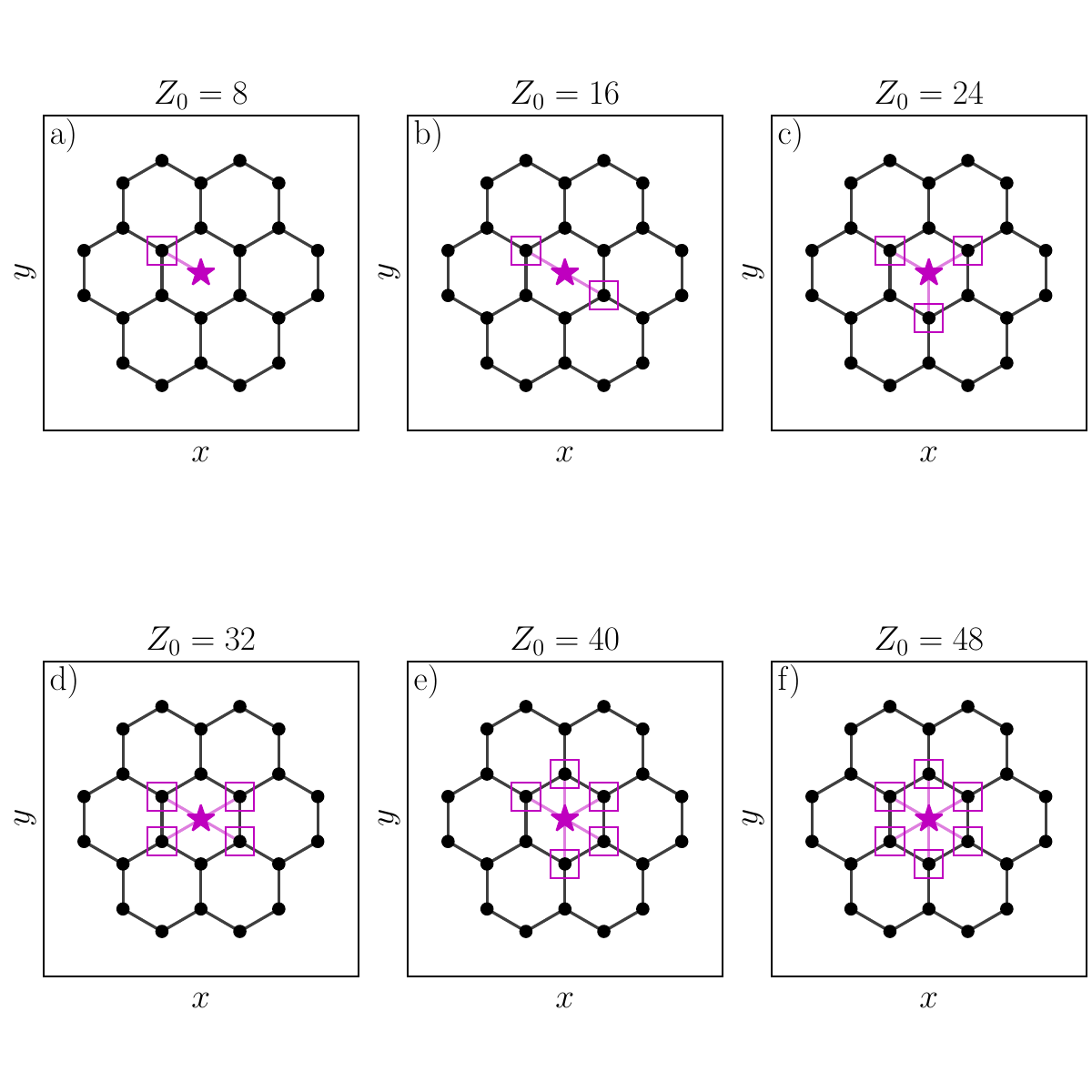}
    \caption{Visualization of the coupling scheme between the honeycomb sites (black dots) and the HCI (purple star). Depending on the initial charge $Z_0$ of the ion, an integer number of $Z_0/8$ sites of the cluster are coupled (purple squares). The spatial layout is chosen such that the coupling is as symmetric as possible.}
    \label{fig:coupling_explained}
\end{figure}

 %
 In the presence of a highly-charged ion (HCI) with initial charge $Z_0$ we use a semiclassical Ehrenfest model according to which the Hamiltonian is modified by an additional diagonal contribution, due to the  Coulomb potential of the ion \cite{balzer_prb16, balzer_cpp_21}
 \begin{align}
     W_{ii}=-\frac{Z_0W_0a_0}{|\boldsymbol{r}_i-\boldsymbol{S}(t)|},
 \end{align}
 with the rescaled Coulomb constant $W_0=e^2/(4\pi\epsilon_0a_0)\approx2.68J_0$. In this paper we focus on simple straight ion trajectories, normal to the cluster plane,  $\boldsymbol{S}(t)= z(t)\boldsymbol{e}_z=(z_0+v_0t)\boldsymbol{e}_z$, without updating the velocity, $v_0$, which is well justified for the comparatively high ion velocities studied below. To reduce the complexity of the model, we neglect off-diagonal contributions. 
 The HCI itself is modeled by a set of degenerate static energy levels with energy $\epsilon$
 \begin{align}
     \hat{H}^{\mathrm{ee}}=\left(-\frac{Z_0W_0}{2}-\epsilon\right)\sum_{i,\sigma}\hat{a}_{i,\sigma}^\dagger\hat{a}_{i,\sigma},
 \end{align}
derived from the value of the Coulomb potential at the point of resonant charge transfer. This implies that electrons on the ion can not decay into lower orbitals, eliminating complex processes, such as Auger recombination or interatomic Coulomb decay~\cite{jahnke_interatomic_2020} from this picture.\\
The charge transfer (ion neutralization) Hamiltonian will be described by a Gaussian amplitude for the transition of electrons from specific system sites into the environment  sites (i.e., ionic orbitals) \cite{balzer_cpp_21}
\begin{align}
    \hat{H}^{\mathrm{se}}=\gamma[\boldsymbol{S}(t)]\sum_{i_1,i_2,\sigma}\delta^{\mathrm{CT}}_{i_1,i_2}\left\{\hat{c}_{i_1,\sigma}^\dagger\hat{a}_{i_2,\sigma}+\hat{a}^\dagger_{i_2,\sigma}\hat{c}_{i_1,\sigma}\right\}\,,\quad\gamma[\boldsymbol{S}(t)]=\gamma_0\exp\left(-\frac{(z(t)+z_{\mathrm{res}})^2}{2d_{\mathrm{w}}^2}\right).
\end{align}
The peak of the Gaussian is located at the point of resonant charge transfer, $z_{\mathrm{res}}=-\sqrt{3}a_0$, has a width of $d_{\mathrm{w}}=0.6a_0$ and an amplitude of $\gamma_0=2.12$, that was determined in Ref.~\cite{balzer_cpp_21}. The charge transfer Kronecker delta, $\delta^{\rm CT}$, determines the coupling scheme between the system and the HCI. In order to mimic realistic charge transfer, the sites nearest to the HCI penetration point are coupled to the ion, and since each site can contribute 2 electrons per each of the four bands, the number of coupled sites is chosen to be $Z_0/8$ (integer). This allows for full neutralization of the ion, while limiting the charge flux (see Fig.~\ref{fig:coupling_explained}). 

\begin{figure}[h!]
    \centering
    \includegraphics[width=0.7\textwidth]{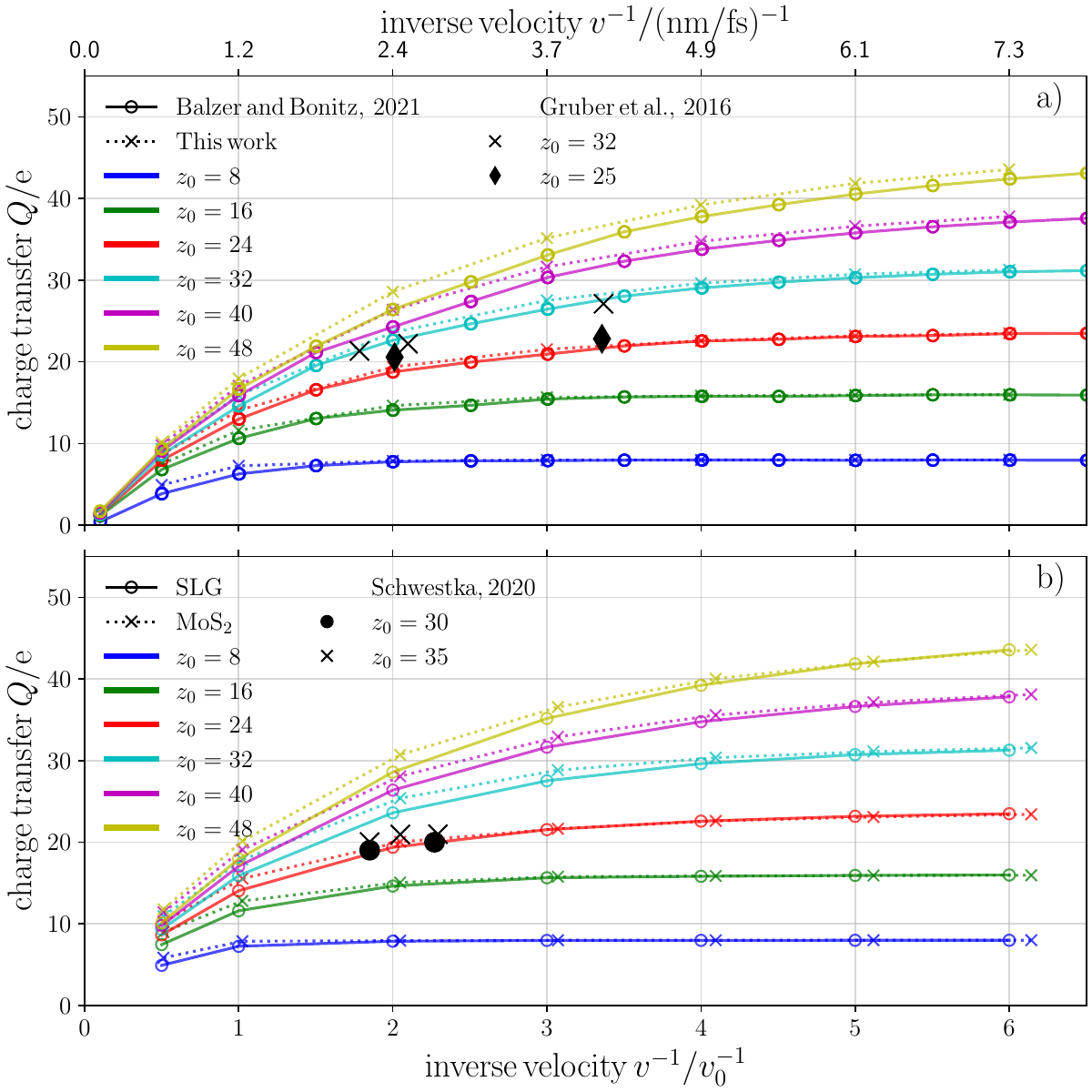}
    \caption{Total charge transfer as a function of the inverse ion velocity for different charge states. a) SLG: comparison between NEGF~\cite{balzer_prb_23} and the present time-local embedding scheme, both in HF approximation, as well as experimental data~\cite{gruber_ultrafast_2016}. b) Comparison between the present simulations for SLG and MoS$_2$, compared to experimental data for MoS$_2$~\cite{schwestka_charge_2020}. }
    \label{fig:velocity_dependency_7}
\end{figure}

An observable which is experimentally accessible is the total charge transferred from the cluster to the HCI
\begin{align}
    Q=4\big\{N_\textup{s}-\braket{N}(t\rightarrow\infty)\big\},
\end{align}
serving as a measure of the intensity of the interaction between the cluster and the ion. The transferred charge is analyzed in detail in Fig.~\ref{fig:velocity_dependency_7} as a function of the inverse ion velocity for various initial charges $Z_0$. The upper figure contains experimental data for single-layer graphene reported by Gruber \textit{et al.}~\cite{gruber_ultrafast_2016} for which the ion velocities are typically low, in the range of $0.1 \dots 0.5 \cdot 10^6$ ms$^{-1} = 0.1 \dots 0.5$  nm/fs which translates into interaction times of $2\dots 10$fs \cite{gruber_ultrafast_2016}. In Fig.~\ref{fig:velocity_dependency_7}.a we reproduce data for two ion charges, $Z_0=25$ and $Z_0=32$, for inverse velocities between 2 and 4 (nm/fs)$^{-1}$ and interaction times between 2 and 4 fs. These data were used to adjust the charge transfer intensity of the model introduced in Ref.~\cite{balzer_cpp_21} that was applied also in the simulations of Ref.~\cite{niggas_prl_22}.
The charge transfer model results of Ref.~\cite{balzer_cpp_21} are shown in the upper figure by the lines with open circles, for six different initial ion charges. The results for the charge transfer of the present work that are obtained from our correlated simulations and a time-local embedding scheme are shown by the lines of the same color, but with crosses. We observe overall very good agreement of the two simulations for graphene, for all initial charges. The largest deviations are observed for the largest initial charges, $Z_0=40, 48$, and low velocities where our improved model predicts up to 7$\%$ higher transferred charge for graphene.

Let us now compare the charge transferred to the ion between the two materials, single-layer graphene (SLG) and MoS$_2$, cf. Fig.~\ref{fig:velocity_dependency_7}, b. Our results are depicted by the colored lines with open circles (SLG) and crosses (MoS$_2$) and predict a systematically slightly higher charge transfer for the case of MoS$_2$ which is in agreement with the experimental observations of Ref.~\cite{schwestka_charge_2020}. There it was noted that, an important reason for the increased charge transfer is the larger layer width of MoS$_2$ leading to a larger interaction time with the ion. In our model no finite monolayer width effects are included and the increased charge transfer of MoS$_2$ results from its larger lattice constant compared to SLG. This will be discussed in more detail below, in the context of Fig.~\ref{fig:density_profiles_5}.
Compared to the experiments our data for the charge transfer appear to be slightly higher, so we also repeated simulations with a slightly reduced charge transfer amplitude $\gamma_0$ compared to the one determined in Ref.~\cite{balzer_cpp_21}, however, this did not have a noticable effect on the electron dynamics in the monolayer. Therefore, in all simulations of this paper we used the same amplitude $\gamma_0=2.12J_0$, for both materials.

\section{Results}\label{s:results}
Using the Hamiltonian as introduced above we have performed extensive simulations of the electron dynamics in the monoloayer, coupled to the ion via Coulomb interaction and the charge transfer Hamiltonian. Our main interest is to study the effect of correlations in the monolayer which were neglected in previous simulations.
 
\subsection{Ion-induced electron dynamics in the target}
\label{ss:e-dynamics}
\begin{figure}[h!]
    \centering
    \includegraphics[width=0.75\linewidth]{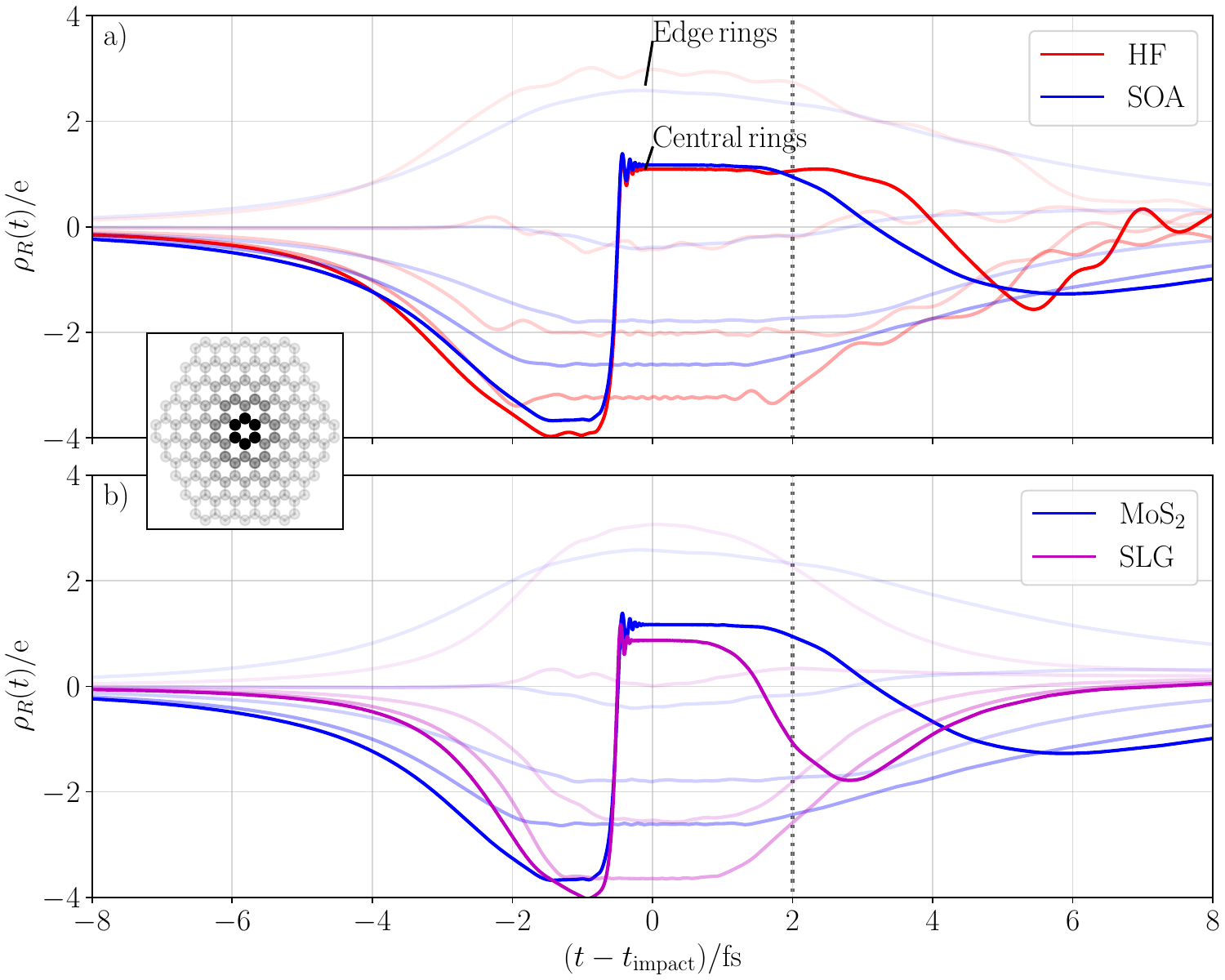}
    \caption{Average charge density on the different rings around the ion impact point, for a) MoS$_2$ comparing HF and SOA self-energies; b) comparing SLG and MoS$_2$ in SOA. The inset depicts the spatial structure of the cluster with nearest-neighbor connections. The brightness of the colors decreases outward. The dotted vertical line indicates the time for which the induced potential is shown in Figs.~\ref{fig:electrostatic_potential_combined} and \ref{fig:potential-3d}.}

    \label{fig:density_profiles_5}
\end{figure}

 To quantify the electron dynamics in the cluster during the interaction with the HCI, we compute the average deviation from charge neutrality on the honeycomb rings $R$
 \begin{align}
     \rho_R(t)=\frac{4e}{L_R}\left(1-i\hbar\sum_{i\in R,\sigma}G_{ii}^{\textup{ss},<}\right),
 \end{align}
 where $R$ is the set of $L_R$ sites which compose a ring. During the approach of the HCI, the central rings accumulate negative charge due to the attractive Coulomb potential of the ion. Subsequently, the outer rings get depleted, resulting in a positive charge. The inclusion of correlations has a significant effect on the charge distribution in the MoS$_2$ cluster: the charge transfer to the ion is reduced and electron migration between the rings is smoothened, see Fig.~\ref{fig:density_profiles_5}, a). At the point of resonant charge transfer, electrons are transferred from the cluster onto the HCI. This transfer is very rapid (sub-femtosecond) in both approximations. During the interaction time with the HCI, a period of constant, positive charge in the center of the cluster can be observed, during which the HCI partially neutralizes. Once the interaction with the ion is over (positive times), electrons flow back into the center of the cluster from outer layers as a plasmon-like wave. Interestingly, the onset of this flow is shifted to later times (more than 1fs) in the HF approximation compared to SOA. The reason is that, in SOA, the change of the charge on the rings is significantly reduced compared to HF (except for the second ring). 
 Also note the high-frequency oscillations of the charge density on the rings for $t\gtrsim 2$fs in the HF approximation, which are dampened out by the inclusion of correlations.\\

 Consider now the differences in the behavior of the two materials, in response to the same perturbation, cf. Fig.~\ref{fig:density_profiles_5}, b). Interestingly, in SLG, the initial charge accumulation initiated by the ion is delayed compared to MoS$_2$, because of the 
 smaller interatomic distance, resulting in a larger, outward facing pressure (electron-electron forces). For the same reason, in graphene, the return to a uniform charge distribution (in particular, radial electron flow towards the center, to compensate for the charge transfer to the ion, sets in significantly earlier than in MoS$_2$, and already around $6~$fs after the ion impact the charges have nearly equilibrated. On the other hand, in MoS$_2$ this process lasts significantly longer.


\subsection{Ion-induced energy dynamics in the monolayer}
\label{ss:energy-dynamics}
From the ion impact-induced time evolution of the Green functions we can compute additional observables. An example of interest are the various contributions to the energy of the electrons in the monolayer, including their kinetic, mean field (Hartree-Fock), and correlation energy, which are computed according to
\begin{figure}[h]
    \centering
    \includegraphics[width=0.7\textwidth]{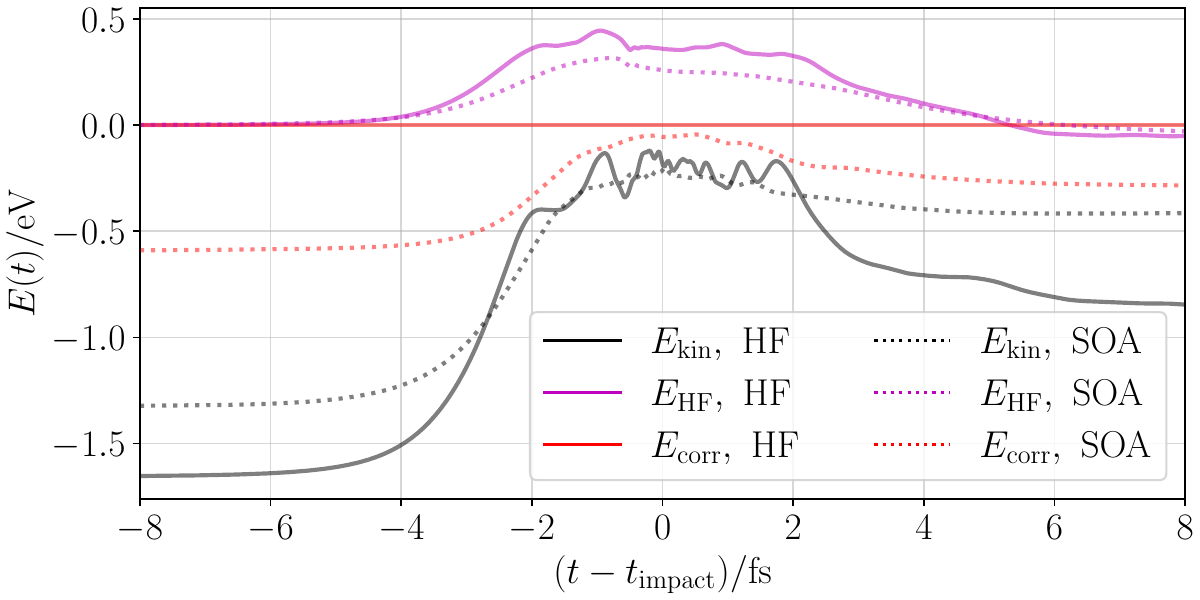}
    \caption{Different energy profiles over the course of the simulation of a $N\U{\mathrm{s}}=150$ MoS$_2$ cluster, interacting with a Xe$^{+32}$ ion with a kinetic energy of $E\U{\mathrm{ion}}=113\,\text{keV}$. Solid lines represent the results of a HF simulation while dotted lines represent the results of a SOA simulation. The different energy forms are kinetic energy $E\U{\mathrm{kin}}$ (black), HF energy $E\U{\mathrm{HF}}$ (purple) and correlation energy $E\U{\mathrm{corr}}$ (blue).}
    \label{fig:energy_profiles_4}
\end{figure}
\begin{align}
    E_{\mathrm{kin}}(t)=i\hbar\mathrm{Tr}\big[\boldsymbol{h}^{\textup{ss},0}\boldsymbol{G}^{\textup{ss},<}(t)\big],\quad E_{\mathrm{HF}}(t)=\mathrm{Tr}\big[\boldsymbol{h}^{\textup{ss},\textup{HF}}(t)\boldsymbol{G}^{\textup{ss},<}(t)\big],\quad E_{\mathrm{corr}}(t)=U(i\hbar)^2\sum_{i}\mathcal{G}_{iiii}^{\uparrow\downarrow\uparrow\downarrow}(t).
\end{align}
Since the system starts in a half-filled configuration, initially the HF-energy is zero. During the approach of the HCI to the monolayer, its Coulomb potential deforms the potential landscape in the center of the cluster, leading to an increase of the kinetic and HF-energies of the electrons. After the HCI removes electrons from the system, the cluster is no longer half-filled and, therefore, exhibits a non-vanishing HF-energy, as well as an increased residual, kinetic energy (see Fig.~\ref{fig:energy_profiles_4}). If correlations are neglected (cf. curves denoted ``HF''), kinetic and mean field energy show strong oscillations. These are however completely damped when correlations are taken into account (curves denoted ``SOA'').
Also note that, in SOA the initial state is correlated and, therefore, carries a non-vanishing (negative) correlation energy (red curve). After the interaction with the HCI, this correlation energy (its magnitude) decreases. We will see later in Sec.~\ref{ss:dos-dynamics} how this change in correlation energy manifests in a change to the DOS of the cluster.

\subsection{Ion-induced doublon dynamics in the monolayer}
\label{ss:doublon-dynamics}
Another quantity that is sensitive to electron-electron interactions is the number of doublons, pairs of electrons with opposite spins on the same site.
\begin{figure}[h]
    \centering
    \includegraphics[width=0.7\textwidth]{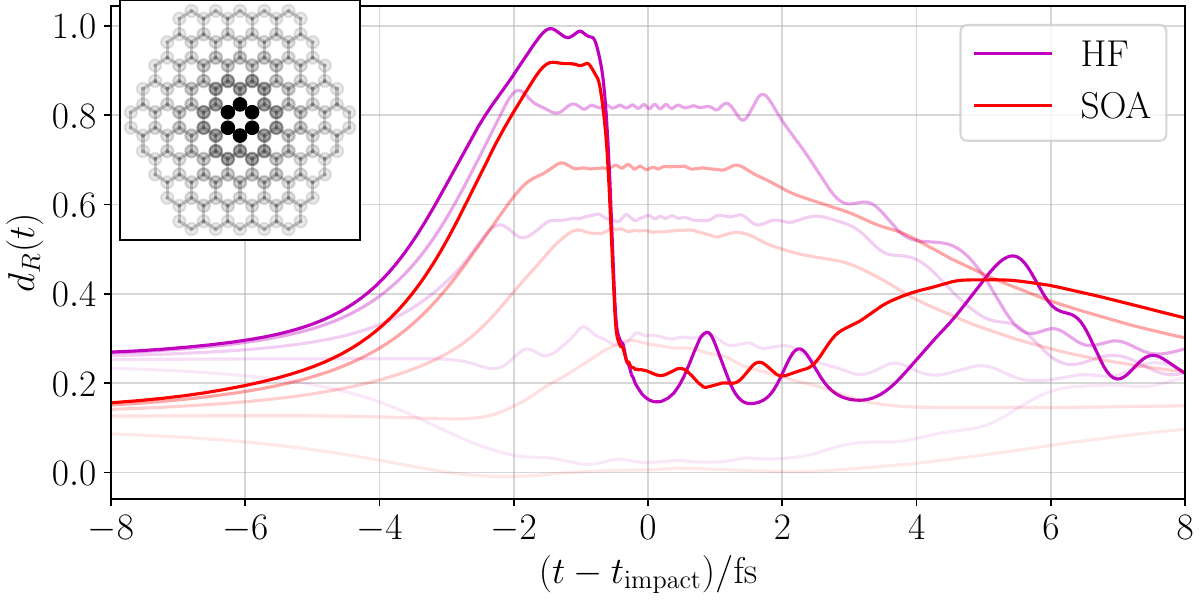}
    \caption{Doublon occupation $d\U{R}(t)$ resolved for the different honeycomb rings during the interaction with the ion. Compared are the time series for MoS$_2$ in HF (purple solid lines) and SOA (red solid lines).}
    \label{fig:doub_occ_1}
\end{figure}
This quantity is important for the charge transport in correlated Fermi systems. For example, after a confinement quench in fermionic atoms doublons and single electrons have been found to diffuse with different speeds \cite{schneider_np_12,schluenzen_prb16}.
It is interesting to spatially resolve the doublon number. On a hexagonal ring with index $R$ their number is given by
\begin{align}
    d_R(t)=\sum_{i\in R}G_{ii}^{\textup{ss},<}(t)G_{ii}^{\textup{ss},<}(t)+\mathcal{G}_{iiii}(t).
\end{align}
Driving a strongly-correlated system out of equilibrium can lead to the formation of a significant number of doublons, which was observed for the impact of a weakly charged ion or multiple ions \cite{balzer_prl_18,borkowski_pss_22}. An even stronger effect is observed in the present excitation scenario with highly-charged ions: during the approach of the HCI the doublon occupation in the central rings increases sharply because of the attractive Coulomb potential of the ion that acts like a time-dependent confinement in the monolayer. Unlike in Ref.~\cite{balzer_prl_18}, this increased occupation does not decay monotonically towards a stable configuration. This is because electrons from the central ring are transferred to the HCI which leads to a rapid decrease of the doublon number in the vicinity of the ion impact point. 
After the interaction time between the HCI and the cluster (after about 2.5 fs), the doublon occupation on the innermost ring increases again and subsequently slowly decays towards the initial value before the ion impact (see Fig.~\ref{fig:doub_occ_1}). In contrast, in HF, this relaxation is much more violent and oscillatory representing artificial effects due to the neglect of correlations.  


\subsection{Induced electrostatic potential}\label{ss:induced-pot}
\begin{figure}[h]
    \centering
    \includegraphics[width=1\textwidth]{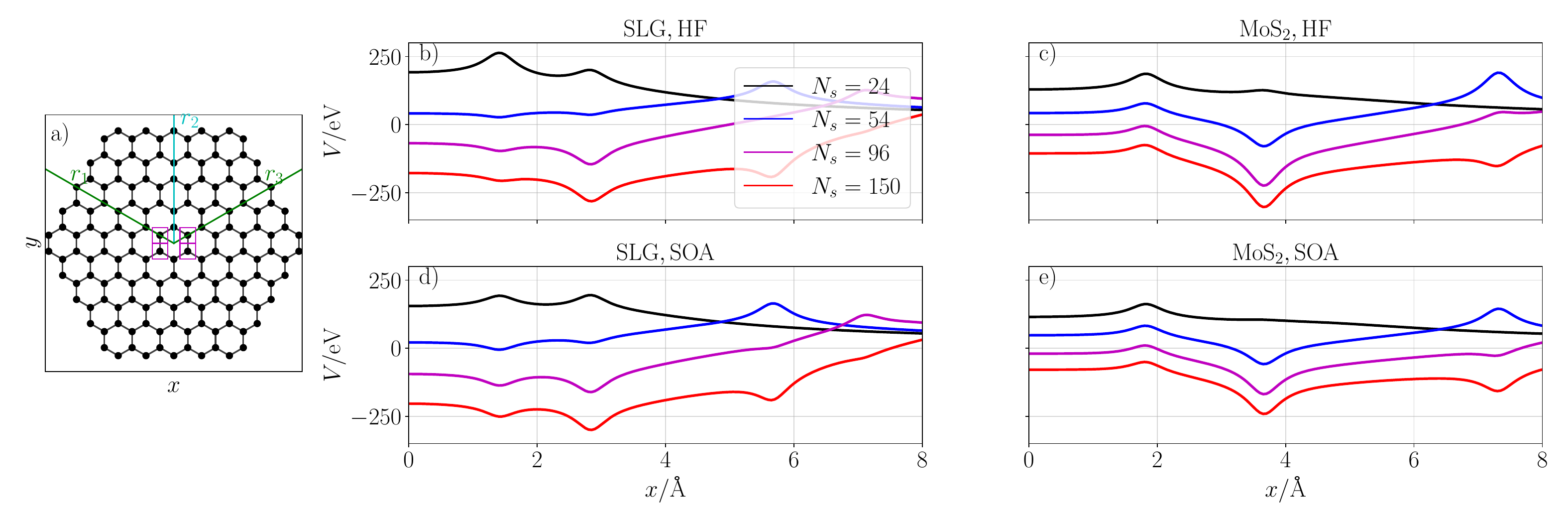}
    \caption{a) Visualization of the calculation of the electrostatic potential according to Eq.~\eqref{eq:electrostatic_potential}. The directions $\boldsymbol{r}\U{1}$ and $\boldsymbol{r}\U{3}$ (green) correspond to paths that involve a site from which resonant charge transfer occurs, while $\boldsymbol{r}\U{2}$ (cyan) corresponds to a path including a site that is not coupled to the ion. The potential is then averaged over all three directions, b)-e) Electrostatic potentials for $t=2$fs, for different simulations (SLG, MoS$_2$ and HF, SOA) and four different system sizes $N_s$ (see legend). The potentials are calculated with Eq.~(\ref{eq:electrostatic_potential}) and the averaging scheme is displayed in a). The behavior at different heights $z$ is shown in Fig.~\ref{fig:potential-3d}.}
    \label{fig:electrostatic_potential_combined}
\end{figure}

An important quantity that was found to crucially influence the emission of secondary electrons after ion impact is the electrostatic potential induced in the cluster as a consequence of the ion Coulomb force and the resonant charge transfer \cite{niggas_prl_22}. In this reference it was observed that this potential is primarily responsible for the dramatic differences in the number of secondary electrons (more than a factor 8 higher in SLG) emitted from SLG and MoS$_2$, respectively.
The space and time-dependent potential follows from the solution of Poisson's equation containing the time-dependent charge density $\rho_i(t)$ in the cluster, where $i$ labels the lattice sites, 
\begin{align}\label{eq:electrostatic_potential}
    V(\boldsymbol{r},t) &=\frac{1}{4\pi\epsilon_0}\sum_i\frac{\rho_i(t)}{|\boldsymbol{r}-\boldsymbol{r}_i|}\,,\\
    \textbf{F}^{\rm SE}(\boldsymbol{r},t) &= |e|\frac{\partial}{\partial \textbf{r}}V(\boldsymbol{r},t)\,,\label{eq:force-on-see}
\end{align}
and we also introduced the force that is produced by the charge density and acts on a secondary electron created at time $t$ and point $\textbf{r}$.
Recall that, due to the 8-fold degeneracy of the Hubbard band, we had to select certain sites on the innermost shell (adjacent to the ion impact point) from which resonant charge transfer to the ion proceeds, cf. Fig.~\ref{fig:coupling_explained}. However, in reality the charge transfer proceeds in a radially symmetric way. To restore this symmetry we average over the different radial directions, as is explained in Fig.~\ref{fig:electrostatic_potential_combined}.a).

\begin{figure}[h!]
    \centering
    \includegraphics[width=1.08\linewidth]{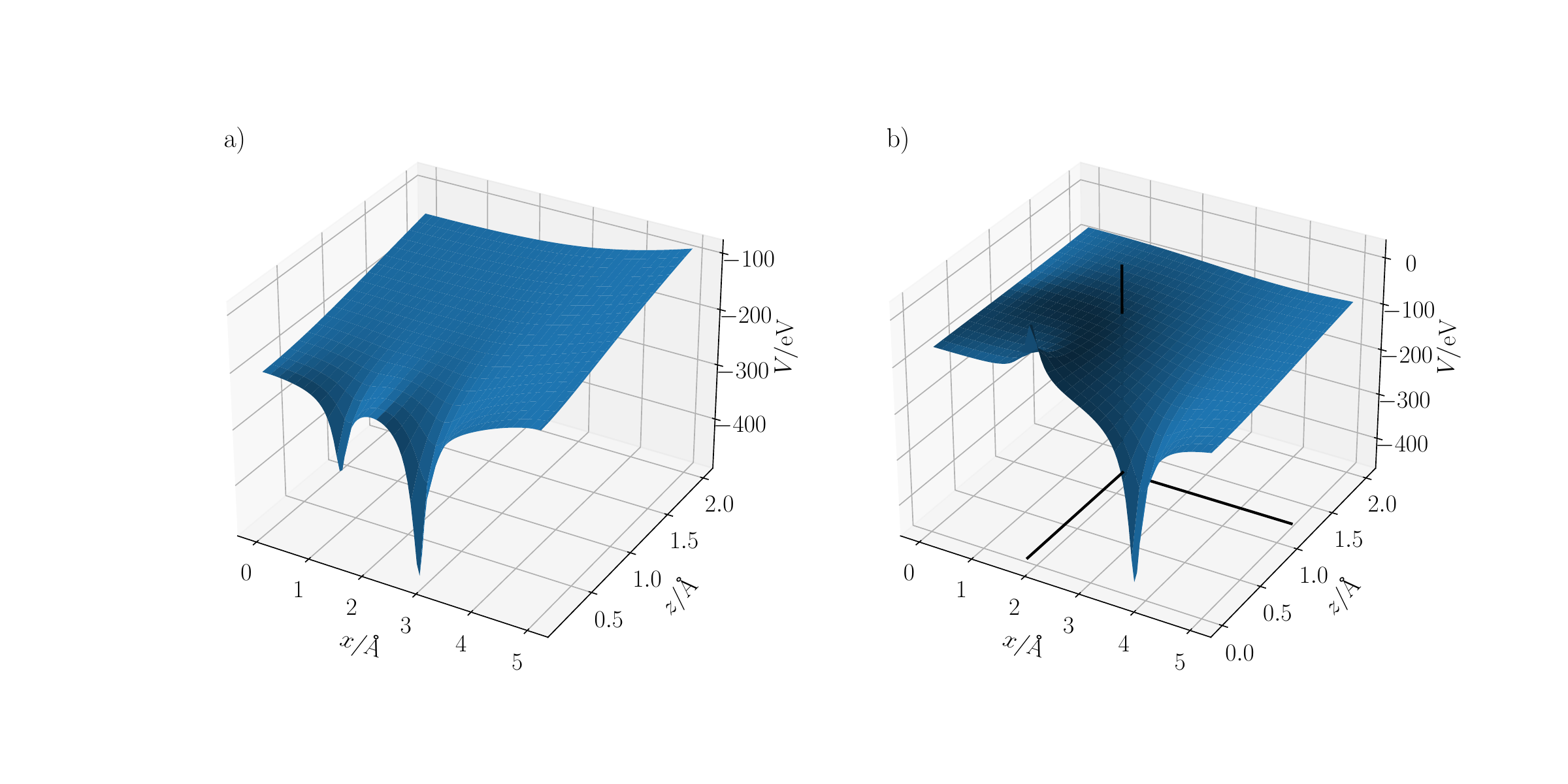}
    \caption{Induced electrostatic potential in the vicinity of a) a SLG and b) a MoS$_2$ monolayer, $2$fs after the impact of a Xe$^{+32}$ ion with initial kinetic energy $E_{\rm ion}=113$keV. The local extrema are located at the hexagonal shells. For SLG (MoS$_2$) there is a local minimum (maximum) at the innermost shell.
    Secondary electrons are generated via ICD in the vicinity of the innermost shell and, for MoS$_2$, experience an attractive force, Eq.~\eqref{eq:force-on-see}. For $z\ge 1.2 ${\AA} (indicated by the black vertical line) the potential increases monotonically to zero and the force on an electron is repulsive. $z$ denotes the height above the layer and $x$ the radial direction, the impact point is the origin. 
    }
    \label{fig:potential-3d}
\end{figure}

A snapshot of the induced potential at time $t=2$fs and at a height $z=0.25${\AA} above the monolayer is shown in Figs.~~\ref{fig:electrostatic_potential_combined}.b)-e), where the corresponding instantaneous charge densities can be read-off from Fig.~\ref{fig:density_profiles_5}, cf. the vertical dotted line. These parameters are chosen because they allow us to directly compare to the HF results that were presented in Fig. 2 of Ref.~\cite{niggas_prl_22}.
First, a strongly non-monotonic behavior in radial direction $x$ is evident where the extrema correspond to the mean shell radii. The innermost peak is located at the shell from which electrons are transferred to the ion. Notice the striking differences in figure parts b) and c) corresponding to SLG and MoS$_2$, respectively: For SLG this innermost peak is a local minimum whereas, for MoS$_2$, it is a local maximum. This has dramatic consequences for secondary electron emission \cite{niggas_prl_22}: Electrons that are generated via the interatomic Coulombic decay (ICD) are subject to the potential $V$ and the associated force on secondary electrons (SE), $\textbf{F}^{\rm SE}$, Eq.~\eqref{eq:force-on-see}. Due to the $r^{-6}$ dependence on the distance between ion and emitting lattice site, ICD generation of secondary electrons will only be effective from the innermost ring. Thus, an ICD generated electron will be able to leave the monolayer towards the detector if the $z$-component of the force is positive (repulsive). Since the potential $V$ overall, decays with increasing height $z$, the $z-$component of the force is typically positive. However, close to the monolayer deviations from this trend occur: While, in case of SLG, the local minimum at the inner ring increases the repulsive force, the local maximum, in case of MoS$_2$, in contrast, gives rise to an attractive force, in the vicinity of the monolayer, see Fig.~\ref{fig:potential-3d}. This explains the much larger number of secondary electrons that are emitted from SLG as compared to MoS$_2$~\cite{niggas_prl_22}.

With our new and improved simulations of the dynamics of the electrons in the monolayer, we can trace the origin of this effect in more detail. In fact, the origin of the local maximum (minimum) of $V$ on the innermost shell can be understood from the density dynamics, cf. Fig.~\ref{fig:density_profiles_5}. In SLG, at $t=2$fs, the electron deficiency (i.e., the positive charge) at the inner ring (due to resonant transfer to the ion) has already been compensated by the flux of electrons from outer rings and has reached a value of $\rho_R \approx -1.5 e$. In contrast, in MoS$_2$, the density at this time is still positive, $\rho_R \approx 1.0 e$. Obviously, this is a consequence of the significantly lower electron mobility in MoS$_2$, as compared to SLG. Let us now discuss the effect of electron-electron correlations on ICD-induced secondary electron emission (SEE) in MoS$_2$, Figs.~\ref{fig:electrostatic_potential_combined} c) and e). 

The simulations with HF and SOA self-energies show very little difference for the height of the potential maximum. The reason can be understood from the electron density on the innermost ring, cf. Fig.~\ref{fig:density_profiles_5}: at time $t=2$fs, SOA and HF simulations are in full agreement with each other. However, at later times, $t\in [2,5]$fs, correlations lead to a more rapid occupation of the inner shell than HF, and already after 3 femtoseconds the density is negative and, therefore, the local maximum of the potential should vanish. Thus, it can be expected that, for slow projectiles, where the interaction period between ion and monolayer increases, the number of secondary electrons may increase significantly. Interestingly, in SLG [cf. Figs.~\ref{fig:electrostatic_potential_combined} b) and d)], correlations in the material give rise to a deepening of the potential minimum and therefore to an increase of the repulsive force. Therefore, SEE should be slightly increased.

Finally, Fig.~\ref{fig:electrostatic_potential_combined} allows us to predict an interesting dependence of SEE on the cluster size, cf. the four lines of different color which represent clusters with $N_s=24$ to $N_s=150$ sites. In MoS$_2$, the maximum on the inner shell does almost not change with cluster size. The reason is that only electrons from the second ring have time to reduce the electron depletion on the innermost ring. In contrast, in SLG the potential minimum becomes slightly deeper when the cluster size increases. A very interesting observation can be made for very small clusters, $N_s \le 54$, for both materials. Here, the potential at the innermost ring is positive, producing an attractive force on secondary electrons which would suppress SEE, even in SLG. The reason is, of course, that for such small clusters the loss of electrons to the ion cannot be compensated at all, so the cluster remains positively charged even in its central part. Therefore, in small clusters, SEE will only be possible for electrons that are created with a sufficiently high kinetic energy so they can overcome the local potential barrier.

\begin{figure}
    \centering
    \includegraphics[width=0.99\linewidth]{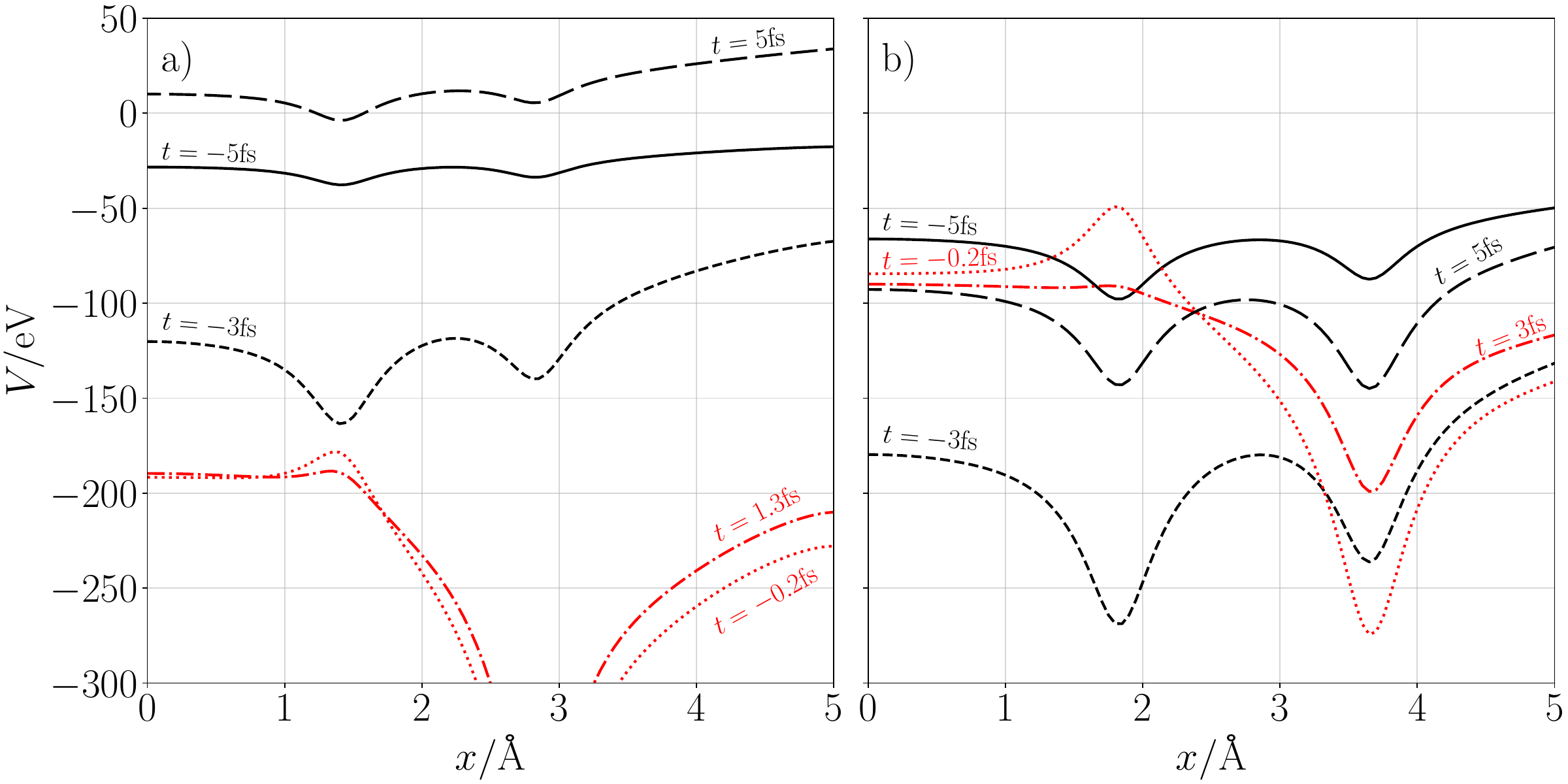}
    \caption{Time evolution of the potential  induced in a) a SLG and b) a MoS$_2$ monolayer by a 32-fold charged ion, at the height $z=0.25 ${\AA} above the layer. When the ion approaches, this potential rapidly develops minima that deepen until the impact, at $t=0$. Red curves show the effect of the resonant charge transfer to the ion from the innermost layer -- the transient formation of a local maximum, see also Fig.~\ref{fig:potential-3d}. The times in panel b) are the same as in Fig.~\ref{fig:dm_8}.}
    \label{fig:mos2-pot-vs-t}
\end{figure}

We conclude this section by presenting, as an overview, the entire time evolution of the induced potential for the cases of SLG and MoS$_2$ in Fig.~\ref{fig:mos2-pot-vs-t}. The plots show the dramatic influence of the Coulomb potential of the ion on the electrons in the two innermost shells. The overall trend is clearly seen on the second shell: The potential starts from a value $V=0$, without the ion, and then rapidly becomes more negative due to accumulation of electrons in the vicinity of the impact point. After reaching the deepest value at $t=0$, the minimum becomes more shallow and eventually vanishes, for $t\to \infty$. 
Now, let us consider the evolution of the potential at the innermost honeycomb. It also rapidly decreases until about $t=-1$fs. However, then charge transfer sets in and removes electrons. As a consequence, at $t=-0.2$fs, the potential rapidly increases and, in the case of MoS$_2$, even exceeds its value at $t=-5$fs. The formation of the local maximum that exists for a transient time of about 3.5 fs in MoS$_2$ and 2 fs in SLG (cf. Fig.~\ref{fig:potential-3d}) is evident.
Note that the behavior at time $t$ is significantly different from $-t$, as a consequence of energy dissipation [cf. Fig.~\ref{fig:energy_profiles_4}] and correlation effects inside the monolayer.

\subsection{Ion-induced evolution of the density of states}\label{ss:dos-dynamics}
Before we can investigate the electron energy spectra of the different simulations, we first have to discuss how to obtain them from the nonequilibrium simulation data. The main problem in time-diagonal methods, such as the G1--G2 scheme, is that spectral information about correlation effects is lost when using HF-propagators for the retarded and advanced Green function in the GKBA, cf. Eq.~(\ref{eq:g1_reconstruction}). In contrast, in two-time approaches, such as real-time NEGF theory \cite{keldysh64}, one has direct access to the spectra of a system. 
Then the time-resolved DOS can be calculated from the spectral function according to~\cite{joost_phd_2022}
 \begin{align*}
	 A\U{ij}\Ho{<}(\omega,T) & =-i\hbar\int\mathrm{d}t\mathrm{d}t^\prime\,\mathcal{S}(t-T)\mathcal{S}(t^\prime-T)e^{-i\omega(t-t^\prime)}G\U{ij}\Ho{<}(t,t^\prime), \\
	 A\U{ij}\Ho{>}(\omega,T) & =i\hbar\int\mathrm{d}t\mathrm{d}t^\prime\,\mathcal{S}(t-T)\mathcal{S}(t^\prime-T)e^{-i\omega(t-t^\prime)}G\U{ij}\Ho{>}(t,t^\prime),  \\
	 \mathrm{DOS}(\omega,T)  & =\sum\U{i}\bigg(A\U{ii}\Ho{>}(\omega,T)+A\U{ii}\Ho{<}(\omega,T)\bigg),
 \end{align*}
 where $\mathcal{S}$ is an artificial broadening, meant to resemble the spectral width of a probe pulse in a pump-probe experiment. However, since, in the G1--G2 scheme, one does not have access to these two-time quantities, one has to approximate the spectrum using only the time-diagonal quantities that are available in the simulation. \\
 One such method is given by Koopmans' theorem~\cite{koopmans_uber_1934}, which states that the ionization energies of a quantum mechanical system can be approximated by the eigenvalues of the single-particle Hamiltonian. 
 Of course, in correlated systems, this holds only approximately and we can only expect qualitative results.
 Koopmans' theorem is well suited for our situation, since the single-particle HF-Hamiltonian can be directly calculated from the one-particle density matrix. To do so, we diagonalize the single-particle Hamiltonian
 \begin{align*}
	 H\Ho{ss,(1)}\U{ij}(t)=\big(\epsilon-W\U{ii}\big[\bs{S}(t)\big]\big)\delta\U{ij}-J\delta\U{\braket{i,j}}+Un\U{ii}(t)\delta\U{ij},
 \end{align*}
 to obtain the eigenvalues $\{E\U{i}\}$, as well as the corresponding eigenstates $\ket{\phi\U{i}}$. The time-dependent DOS is then given by
 \begin{align*}
   \mathrm{DOS}(E,t)=\sum\U{i}\mathcal{S}[E-E\U{i}(t)],\quad \mathcal{S}(E)=\frac{1}{2\sigma^2}e^{-\frac{E^2}{\sigma^2}},
 \end{align*}
 with a spectral width of $\sigma=0.4J_0$. To calculate the ionization spectrum we also need the spectral weights
 \begin{align*}
	 p\U{i}(t)=\bra{\phi\U{i}}\hat{n}(t)\ket{\phi\U{i}},
 \end{align*}
 which are the diagonal elements of the single-particle density matrix $\bs{n}$ in the eigenbasis of the single-particle Hamiltonian, giving the occupation of those states. The full spectrum of occupied states is then given by
 \begin{align*}
	 A\Ho{<}(E,t)=\sum\U{i}p\U{i}(t)\mathcal{S}[E-E\U{i}(t)].
 \end{align*}
 \begin{figure}[h!]
    \centering
    \includegraphics[width=0.99\textwidth]{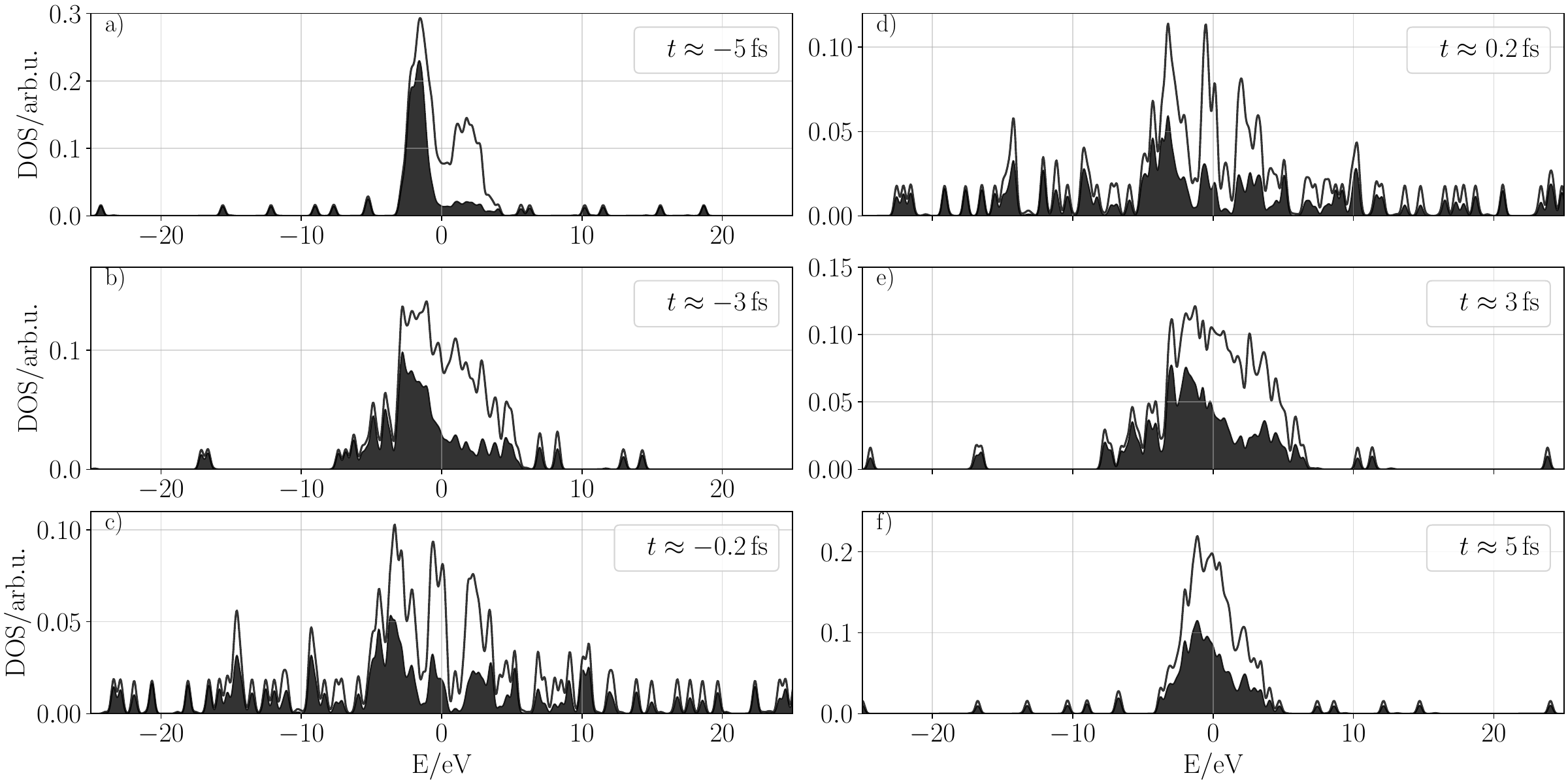}
    \caption{Density of states of a MoS$_2$ monolayer  excited by a Xe$^{+32}$ ion of energy $E_{\rm ion}=113$keV, for 6 times indicated in the figure (same times as in Fig.~\ref{fig:mos2-pot-vs-t}). The ion penetrates the layer at time $t=0$. Filled areas correspond to the occupied states. Note that, already at -5fs the Coulomb force of the ion leads to excitation of electrons above the Fermi level. Results of SOA simulations.}
    \label{fig:dm_8}
\end{figure}

Below, we present simulation results for the ultrafast electron dynamics in a MoS$_2$ monolayer during the impact of a 32-fold charged Xe-ion (impact occurs at time $t=0$). The left (right) part of Fig.~\ref{fig:dos_explained_3.pdf} shows the spectra before (after) the ion impact. The white areas show the entire density of states whereas the filled areas are the occupied states.
Note that, given the energy-time uncertainty, instantaneous nonequilibrium spectra have to be treated with care. Each of the curves has been generated by performing a time average over a small window of width $\Delta t=0.1 t_0$, centered at the time given in the figures.
The initial energy of the ion is $E_{\rm ion}$, corresponding to a velocity of $0.407$nm/fs and an interaction time with the cluster, $\tau\approx \frac{d_{\rm eff}}{v}\approx 2.2$fs where $d_{\rm eff} \approx 9${\AA} \cite{schwestka_charge_2020}.

In Fig.~\ref{fig:dos_explained_3.pdf}.a), at $t=-5$fs, the influence of the ion is still weak, and the DOS is broadened, as a result of correlations that were generated by adiabatic switching, in comparison to the tight-binding DOS of a honeycomb lattice, at half filling. Furthermore, the Coulomb force of the ion has already given rise to acceleration of some of the electrons towards the impact point which leads to occupation of a few high-energy states. In panel b) the influence of the ion is already strong; it deforms and broadens the spectrum. The fraction of electrons excited to higher energies is increased further. In panel c) the ion reaches the minimal distance from the lattice sites which, for the innermost honeycomb, is on the order of the bond length. Correspondingly, electrons on the inner ring experience a very strong attractive potential, in the range of $-40$ to $-200$eV, depending on the charge transfer which has already started. As a result the spectrum becomes very broad, extending to low energies (due to the ion potential) and high energies due to acceleration effects.

After the ion has left the monolayer, its Coulomb force quickly fades and the spectrum starts to relax. Consequently, both, the low energy and high-energy states become only weakly occupied and a main peak remains. It significantly differs from the characteristic two-peak structure that was present before the ion impact, cf. Fig. a) reflecting substantial energy absorption by the electrons in the monolayer. This interpretation is also confirmed by the time evolution of the energy contributions, cf. Fig.~\ref{fig:energy_profiles_4}. There, we observed a significant increase of the correlation energy and an even stronger increase of the kinetic energy that was induced by the projectile.

\begin{figure}[h]
    \centering
    \includegraphics[width=0.7\textwidth]{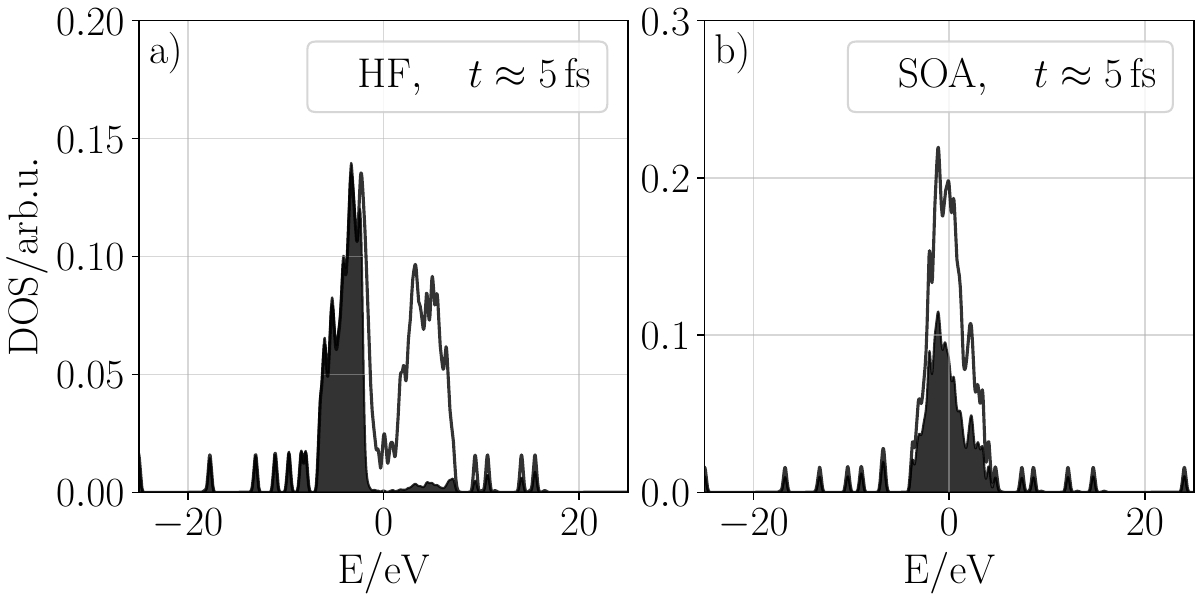}
    \caption{Nonequilibrium density of states of a MoS$_2$ monolayer at $t=5$fs after the ion impact. Comparison of HF (left panel) and SOA (right) simulations. Filled areas correspond to the occupied states. }
    \label{fig:dm_7}
\end{figure}
Finally, we are in a position to investigate the influence of second-order Born correlations on the DOS, in comparison to a Hartree-Fock calculation. This comparison is shown for the final state, $t=5$fs, in Fig.~\ref{fig:dm_7}. Evidently, correlations lead to a merging of the two tight binding bands, as well as to an increased occupation of higher-energy states. Concluding this section, we mention that recently an alternative approach to correlated spectra in G1--G2 simulations was proposed by Reeves and Vlcek \cite{reeves_prl_24} which will be tested in forthcoming work.

 \section{Discussion}\label{s:discussion}
In this paper, we have extended the discussion of Ref.~\cite{niggas_prl_22} about the theoretical description of the interaction of slow highly-charged ions with monolayers of quantum materials. Our simulations are based on a nonequilibrium Green functions description of the electrons in the monolayer within a Hubbard model. The monolayer is coupled to the ion via an embedding scheme that models resonant charge transfer to the ion leading to ultrafast (partial) neutralization. The main improvements compared to Ref.~\cite{niggas_prl_22} are that we included electron-electron correlations via second-order Born self-energies. For this purpose, we applied the G1--G2 scheme \cite{schluenzen_prl_20} in combination with a time-linear embedding scheme \cite{balzer_prb_23}.\\
One of the key results of Ref.~\cite{niggas_prl_22} was the experimental observation of a dramatic difference (about a factor 7) in secondary electron emission, caused by the impact of a highly-charged ion, from a monolayer of graphene as compared to a  MoS$_2$ monolayer. Our theoretical analysis allows us to confirm that this effect is caused by the different structural and electronic properties of the two materials, in particular the smaller lattice spacing and the much higher electron mobility in graphene, as compared to MoS$_2$. The different behavior of the two materials was demonstrated by detailed results for the time and space dependence of the electrostatic potential induced inside of the monolayer, cf. Figs.~\ref{fig:potential-3d} and \ref{fig:mos2-pot-vs-t}. An interesting new result is that the reduction of SEE is confined to fast ionization processes that occur within 3fs after the ion impact. Here, only electrons that are created with sufficiently high kinetic energy are able to leave the material. On the other hand, for slower projectiles, where the interaction duration with the material is longer, e.g., Ref.~\cite{Niggas_phys-com_21}, SEE is predicted to be substantially enhanced, also in MoS$_2$.

The confirmation of the theoretical explanation given in Ref.~\cite{niggas_prl_22} is important since there the analysis was restricted to time-dependent Hartree-Fock simulations, whereas quantum materials, such as MoS$_2$ are characterized by strong electronic correlations.  By including correlations on the second-order Born level we cannot only substantiate the previous results but also directly predict the influence which correlations have on ICD-induced secondary electron emission. As is demonstrated in Fig.~\ref{fig:density_profiles_5}.a), correlations directly affect the speed at which the charge on the inner ring is restored after ion neutralization: while a HF calculation predicts a sign change of the charge on the inner ring around $t=4$fs, with correlations taken into account, this happens already around $t=3$fs. At the same time, second order Born selfenergies provide only the lowest order correlation corrections. Further improvements require an extension to GW selfenergies in order to verify the influence of stronger correlations and of excitonic effects. \\

Due to the high computational effort, the present G1--G2 simulations were restricted to cluster sizes $N_s \le 150$. While finite clusters of quantum materials are an interesting subject in their own right, the comparison with the macroscopic monolayers used in the experiment carries finite size effects. While long simulations are possible, the main bottleneck is the storage of the four-dimensional tensor of the two-particle Green function. We expect that, by using the recently developed quantum fluctuations approach \cite{schroedter_cmp_22,schroedter_23,schroedter_cpp_24}, it will be possible in the near future to significantly extend the system size and also consider multiple layers of quantum materials and study their interaction with highly-charged ions \cite{Niggas_phys-com_21}.\\

The direct link between the SEE yield and the shape of the induced electrostatic potential was demonstrated in Ref.~\cite{niggas_prl_22} by means of semiclassical molecular dynamics simulations for the ICD-generated electrons. This procedure allows one to reproduce the experimentally observed ratio of SEE yields  of SLG and MoS$_2$. At the same time, presently no quantum mechanical results for the states the ICD electrons are ``born'' into and their relative population are available what has required to invoke additional approximations for the initial conditions of the MD simulations \cite{niggas_prl_22}. For this reason the theoretical energy-resolved SEE spectra presently cannot fully explain the measured spectra. To improve on the initial conditions and to take into account the quantum mechanical ICD rates, e.g., \cite{averbukh_prl_04,Niggas_phys-com_21}, remains an important topic for future studies. Another issue of current investigation is the inclusion of nearest neighbor and higher-order interaction  terms, e.g. via the PPP model. On the other hand, it will be interesting to extend the simulations to more bands, e.g.~\cite{shanavas_effective_2015,liu_three-band_2013}, to more accurately capture the electronic structure.
Improvements along these lines will allow one to increase the accuracy of the current predictions for SEE generation in quantum materials, a topic that is also of interest to low-temperature plasma applications \cite{Bonitz_fcse_19}. Our present analysis based on correlated many-body simulations, has clearly shown the strength of our approach: it allows one to systematically study various combinations of quantum materials, as well as initial charge and energy of the projectiles and, ultimately,  to control and optimize the secondary electron emission.


 \section{Acknowledgments}\label{s:acknowledgments}
We thank Anna Niggas and Richard Wilhelm for discussions on details of the highly-charged ion experiments at TU Wien.
This work has been supported in part by the Deutsche Forschungsgemeinschaft via grant BO1366/16.


 \bibliographystyle{pss_title}

\bibliography{thesis-ref,mb-ref}
\end{document}